\documentclass[hidelinks,twoside,twocolumn,9pt]{article}
\usepackage{extsizes}
\usepackage[super,sort&compress,comma]{natbib}
\usepackage[version=3]{mhchem}
\usepackage[left=1.5cm, right=1.5cm, top=1.785cm, bottom=2.0cm]{geometry}
\usepackage{balance}
\usepackage{mathptmx}
\usepackage{sectsty}
\usepackage{graphicx} 
\usepackage{lastpage}
\usepackage[format=plain,justification=justified,singlelinecheck=false,font={stretch=1.125,small,sf},labelfont=bf,labelsep=space]{caption}
\usepackage{float}
\usepackage{fancyhdr}
\usepackage{fnpos}
\usepackage[english]{babel}
\addto{\captionsenglish}{%
  
}
\usepackage{array}
\usepackage{droidsans}
\usepackage{charter}
\usepackage[T1]{fontenc}
\usepackage[usenames,dvipsnames]{xcolor}
\usepackage{setspace}
\usepackage[compact]{titlesec}
\usepackage{hyperref}
\usepackage{epstopdf}%This line makes .eps figures into .pdf - please comment out if not required.

\definecolor{cream}{RGB}{222,217,201}

\usepackage{siunitx}
\usepackage{amsmath,bm}
\usepackage{amssymb}
\usepackage{threeparttable}
\usepackage{color,soul}
\usepackage{lmodern}

\newcommand{\vect}[1]{\boldsymbol{#1}}
\newcommand{\vects}[1]{\boldsymbol{#1}}

\newcommand{\grad}{\vects{\nabla}}

\newcommand{\bkt}[1]{\left[ #1 \right]}
\newcommand{\tsym}[1]{\boldsymbol{#1}} %2nd order tensor symbol
\newcommand{\avg}[1]{\left \langle #1 \right \rangle}

\newcommand{\cfl}{c_0}
\newcommand{\rhofl}{\rho_0}

\newcommand{\bcdot}{\boldsymbol{\cdot}}

\newcommand*{\scaleFactor}{0.75}

\let\Gamma\varGamma
\let\Delta\varDelta
\let\Theta\varTheta
\let\Lambda\varLambda
\let\Xi\varXi
\let\Pi\varPi
\let\Sigma\varSigma
\let\Upsilon\varUpsilon
\let\Phi\varPhi
\let\Psi\varPsi
\let\Omega\varOmega
\begin{document}

\pagestyle{fancy}
\thispagestyle{plain}
\fancypagestyle{plain}{
%%%HEADER%%%
\renewcommand{\headrulewidth}{0pt}
}
%%%END OF HEADER%%%

%%%PAGE SETUP - Please do not change any commands within this section%%%
\makeFNbottom
\makeatletter
\renewcommand\LARGE{\@setfontsize\LARGE{15pt}{17}}
\renewcommand\Large{\@setfontsize\Large{12pt}{14}}
\renewcommand\large{\@setfontsize\large{10pt}{12}}
\renewcommand\footnotesize{\@setfontsize\footnotesize{7pt}{10}}
\makeatother

\renewcommand{\thefootnote}{\fnsymbol{footnote}}
\renewcommand\footnoterule{\vspace*{1pt}% 
\color{cream}\hrule width 3.5in height 0.4pt \color{black}\vspace*{5pt}} 
\setcounter{secnumdepth}{5}

\makeatletter 
\renewcommand\@biblabel[1]{#1}            
\renewcommand\@makefntext[1]% 
{\noindent\makebox[0pt][r]{\@thefnmark\,}#1}
\makeatother 
\renewcommand{\figurename}{\small{Fig.}~}
\sectionfont{\sffamily\Large}
\subsectionfont{\normalsize}
\subsubsectionfont{\bf}
\setstretch{1.125} %In particular, please do not alter this line.
\setlength{\skip\footins}{0.8cm}
\setlength{\footnotesep}{0.25cm}
\setlength{\jot}{10pt}
\titlespacing*{\section}{0pt}{4pt}{4pt}
\titlespacing*{\subsection}{0pt}{15pt}{1pt}
%%%END OF PAGE SETUP%%%

%%%FOOTER%%%
\fancyfoot{}
%\fancyfoot[LO,RE]{\vspace{-7.1pt}\includegraphics[height=9pt]{head_foot/LF}}
%\fancyfoot[CO]{\vspace{-7.1pt}\hspace{13.2cm}\includegraphics{head_foot/RF}}
%\fancyfoot[CE]{\vspace{-7.2pt}\hspace{-14.2cm}\includegraphics{head_foot/RF}}
%\fancyfoot[RO]{\footnotesize{\sffamily{1--\pageref{LastPage} ~\textbar  \hspace{2pt}\thepage}}}
\fancyfoot[RO]{\footnotesize{\sffamily{\thepage}}}
%\fancyfoot[LE]{\footnotesize{\sffamily{\thepage~\textbar\hspace{3.45cm} 1--\pageref{LastPage}}}}
\fancyfoot[LE]{\footnotesize{\sffamily{\thepage}}}
\fancyhead{}
\renewcommand{\headrulewidth}{0pt}
\renewcommand{\footrulewidth}{0pt}
\setlength{\arrayrulewidth}{1pt}
\setlength{\columnsep}{6.5mm}
\setlength\bibsep{1pt}
%%%END OF FOOTER%%%

%%%FIGURE SETUP - please do not change any commands within this section%%%
\makeatletter 
\newlength{\figrulesep} 
\setlength{\figrulesep}{0.5\textfloatsep}

\newcommand{\topfigrule}{\vspace*{-1pt}% 
\noindent{\color{cream}\rule[-\figrulesep]{\columnwidth}{1.5pt}} }

\newcommand{\botfigrule}{\vspace*{-2pt}% 
\noindent{\color{cream}\rule[\figrulesep]{\columnwidth}{1.5pt}} }

\newcommand{\dblfigrule}{\vspace*{-1pt}% 
\noindent{\color{cream}\rule[-\figrulesep]{\textwidth}{1.5pt}} }

\makeatother
%%%END OF FIGURE SETUP%%%

%%%TITLE, AUTHORS AND ABSTRACT%%%
\twocolumn[
  \begin{@twocolumnfalse}
\vspace{1em}
\sffamily
\begin{tabular}{m{13.5cm} p{4.5cm} }

\noindent\LARGE{\textbf{Acoustophoresis of \textit{Legionella} species in water and the influence of collective hydrodynamic focusing}} %& %\includegraphics{head_foot/DOI}
%\\%Article title goes here instead of the text "This is the title"

\vspace{0.1cm} %& \vspace{0.3cm} \\

 \noindent\large{Alen Pavlic,\textit{$^{\ast a}$} Marjan Veljkovic,\textit{$^{b}$} Lars Fieseler,\textit{$^{b}$} and J{\"u}rg Dual\textit{$^{a}$}} %& \\%Author names go here instead of "Full name", etc.

 \vspace{0.6cm}

\noindent\normalsize{\textit{Legionella} are gram-negative, facultative intracellular, and pathogenic bacteria that pose a risk for human health and cause significant energy losses due to extensive preventive heating-up of water installations. We investigate acoustically-driven motion~\textemdash~acoustophoresis~\textemdash~of several \textit{Legionella} species, \textit{Escherichia coli}, \textit{Pseudomonas aeruginosa}, and \textit{Acanthamoeba castellanii}, a common \textit{Legionella} host in water. All the investigated cells can be acoustically manipulated in an ultrasonic standing wave in water, as they possess a non-zero acoustic contrast that is positive for all of the cells, leading to the focusing into pressure nodes of the standing wave.
Multi-body simulations indicate that an increase in cell concentration could significantly accelerate the rate of focusing due to hydrodynamic interactions~\textemdash~a phenomenon that we call \textit{collective hydrodynamic focusing (CHF)}. 
The results form a foundation for acoustic manipulation of bacteria and could pave a path towards acoustically-aided detection of \textit{Legionella} in water.} %& %\includegraphics{head_foot/dates} 
%%\\%The abstrast goes here instead of the text "The abstract should be..."

\end{tabular}

 \end{@twocolumnfalse} \vspace{1.0cm}

    ]
%%%END OF TITLE, AUTHORS AND ABSTRACT%%%

%%%FONT SETUP - please do not change any commands within this section
\renewcommand*\rmdefault{bch}\normalfont\upshape
\rmfamily
\section*{}
\vspace{-1cm}

%%%FOOTNOTES%%%

\footnotetext{\textit{$^{a}$~Institute for Mechanical Systems, Swiss Federal Institute of Technology Zurich, Z{\"u}rich, Switzerland.}}
\footnotetext{\textit{$^{b}$~Institute of Food and Beverage Innovation, Zurich University of Applied Sciences, W{\"a}denswil, Switzerland.}}
\footnotetext{\textit{$^{\ast}$}~\textit{E-mail: apavlic@ethz.ch}}

%Please use \dag to cite the ESI in the main text of the article.
%If you article does not have ESI please remove the the \dag symbol from the title and the footnotetext below.
%\footnotetext{\dag~Electronic Supplementary Information (ESI) available: [details of any supplementary information available should be included here]. See DOI: 00.0000/00000000.}
%additional addresses can be cited as above using the lower-case letters, c, d, e... If all authors are from the same address, no letter is required

%\footnotetext{\ddag~Additional footnotes to the title and authors can be included \textit{e.g.}\ `Present address:' or `These authors contributed equally to this work' as above using the symbols: \ddag, \textsection, and \P. Please place the appropriate symbol next to the author's name and include a \texttt{\textbackslash footnotetext} entry in the the correct place in the list.}

%%%END OF FOOTNOTES%%%

%%%MAIN TEXT%%%%
\clearpage
\section*{Introduction}

\textit{Legionella} is a genus of gram-negative bacteria that are harmful for human health and the causative agent of legionellosis that can appear as legionnaires' disease, Pontiac fever, or Pittsburgh pneumonia.\cite{bartram2007legionella}
Due to the danger of \textit{Legionella}, there are regulations in place to monitor and minimize their presence in water installations.
A common approach to eradicate \textit{Legionella} is periodically heating-up water supplies to above $\SI{60}{\celsius}$. However, periodic heating requires continued energy supply.\cite{van2019overview}
To save energy and to minimize the general impact of diseases originating from \textit{Legionella} bacterial monitoring and analysis could be further improved. Currently, the detection of \textit{Legionella} from water samples relies on manual collection of a rather large sample ($\SI{1}{\liter}$), followed by labour-intensive and time-consuming culturing (up to $10$ days).\cite{chatfield2013culturing} Low concentration limits stipulated by legislature and the associated low abundance of \textit{Legionella} further complicate the process. One way to improve the monitoring of \textit{Legionella} presence could be an up-concentration of \textit{Legionella} in an at-line fashion, prior to the analysis, combined with the use of novel and faster detection systems. Such a combined system could accomplish a sustainable and reliable real-time detection of \textit{Legionella}, leading to considerable energy savings and to the reduction of legionellosis risk on a global scale.
    
The up-concentration of \textit{Legionella} could be accomplished through the means of acoustophoresis,\cite{bruus2011forthcoming} which is known to be a non-destructive method for manipulation of microscopic objects at relatively high throughputs (flow rates of $>\SI{100}{\micro\liter\per\minute}$ per device\cite{undvall2022inertia}). Acoustophoresis is often used in biomedical applications to manipulate cells without affecting their viability, provided that the operating power remains below thresholds for the onset of cavitation and for reaching the temperatures harmful to cells.\cite{apfel1982acoustic,wiklund2012acoustofluidics}
The acoustic field giving rise to acoustophoresis can for example stem from diagnostic or therapeutic medical ultrasound, advanced flow cytometers that use standing waves for higher throughput and accuracy,\cite{goddard2006ultrasonic} or lab-on-a-chip devices for characterization of mechanical properties of eukaryotic cells and organisms.\cite{hartono2011chip,wang2018single,baasch2018acoustic,jimenez2022acoustophoretic}
However, acoustic manipulation of bacteria is limited, mainly due to the small size of bacterial cells that leads to a small acoustic radiation force (ARF), which is the main force in acoustophoresis and scales with the volume of a cell.
The acoustic streaming that is generally present in acoustofluidic devices acts on the cells through the Stokes drag, which scales with the radius of a cell. This difference in scaling of the two competing forces can be characterized with a critical radius, above which the cell motion is dominated by the ARF, while the smaller cells move with the streaming flow. For a cell-like material in water, assuming a standing ultrasonic wave with the frequency $f$ of $\SI{2}{\mega\hertz}$, the critical radius is typically $\approx \SI{0.7}{\micro\meter}$ and scales with the frequency as $\sqrt{1/f}$.\cite{barnkob2012acoustic}
    
Despite the scaling of forces hindering the acoustic manipulation of bacteria like \textit{Legionella}, with the equivalent sphere radius (ESR) in the range of $\sim \SI{0.5}{\micro\meter}$,\cite{rodgers1979ultrastructure} various methods were developed in the recent years. For example, using larger ``seed'' particles that trap bacteria supposedly with acoustic interaction forces,\cite{hammarstrom2012seed,habibi2019sound} using two-dimensional acoustic fields to avoid the disturbance from the acoustic streaming,\cite{antfolk2014focusing} or simply scaling down the dimensions of the acoustofluidic device and increasing the frequency to boost the acoustic radiation force.\cite{ugawa2022reduced} Acoustophoresis of bacteria was so far demonstrated on, for example, \textit{Salmonella} Typhimurium,\cite{schwarz2012ultrasonic} \textit{Escherichia coli},\cite{antfolk2014focusing,gutierrez2018induced,zhao2020disposable} \textit{Enterobacter cloacae},\cite{ugawa2022reduced} and even \textit{Legionella pneumophila},\cite{jepras1989agglutination} but in a fixed state and with potentially dominating influence of agglutination.

The fundamental behaviour of biological cells in a standing acoustic wave was so far characterized for various eukaryotic cells and microorganisms in terms of the acoustic contrast factor $\Phi$ that determines the magnitude of the ARF (ARF $\propto \Phi$) and the locations of stable equilibria of cells in the wave at either the pressure nodes ($\Phi>0$) or pressure antinodes ($\Phi<0$). \cite{hartono2011chip,wang2018single,baasch2018acoustic,jimenez2022acoustophoretic} To the best of our knowledge, there are no such characterizations reported for bacteria, apart from some studies qualitatively and indirectly implying a positive acoustic contrast factor of several bacteria in water for \textit{Salmonella} Typhimurium,\cite{schwarz2012ultrasonic} \textit{Escherichia coli},\cite{antfolk2014focusing,gutierrez2018induced,zhao2020disposable} and \textit{Enterobacter cloacae}.\cite{ugawa2022reduced}

\begin{figure*}[h]
\centering
\includegraphics[scale=\scaleFactor]{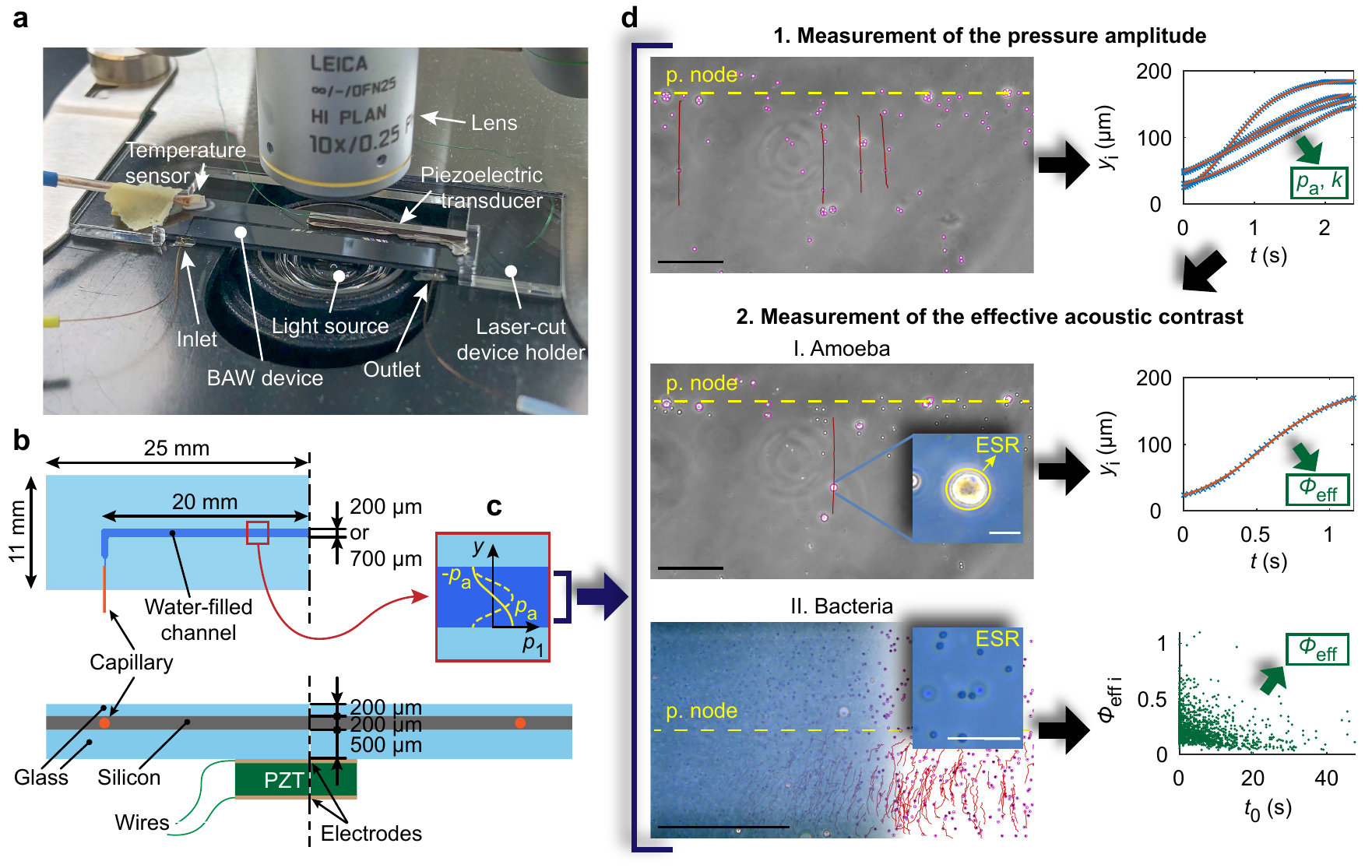}
\caption{Experimental setup and procedure. (a) Description of the measurement setup and the key components. The motion of bacteria, amoebae, and polystyrene particles inside the water-filled channel is observed through a phase contrast microscope from above. (b) The structure and dimensions of the glass-silicon-glass bulk-acoustic-wave (BAW) device, driven by a piezoelectric transducer (PZT). During experiments, PZT is facing downwards, toward the light source. (c) A standing $\lambda/2$-wave (solid line) or a $\lambda$-wave (dashed line) with a pressure amplitude $p_{\mathrm{a}}$ is established in the water-filled channel across its width. The channel width in experiments is either $\SI{200}{\micro\meter}$ or $\SI{700}{\micro\meter}$. (d) The experiments of a given experimental series rely on the measurement of $p_{\mathrm{a}}$ and wavenumber $k$ by tracking the motion of polystyrene particles of known size and properties. Measured $p_{\mathrm{a}}$, $k$, and the equivalent sphere radius (ESR) from high-magnification images (1000x) then serve as an input parameter for the determination of the effective acoustic contrast factor $\Phi_{\mathrm{eff}}$, based on a single-cell tracking. In the case of bacteria, experimental videos are segmented to aid the tracking; many individual bacteria are tracked across several experiments, yielding an average $\Phi_{\mathrm{eff}}$ of a sample. White scale bar corresponds to $\SI{10}{\micro\meter}$, black scale bar corresponds to $\SI{100}{\micro\meter}$.
}
\label{fig:device_setup}
\end{figure*}

We study acoustophoresis of several species of \textit{Legionella}, \textit{E. coli}, and \textit{P. aeruginosa} that are relevant for water safety.\cite{bartram2007legionella,world2017guidelines} We experimentally determine the acoustic contrast factor of 10 \textit{Legionella} species, specifically \textit{L. pneumophila}, \textit{L. fraseri}, \textit{L. pascullei}, \textit{L. anisa}, \textit{L. cherrii}, \textit{L. micdadei},
\textit{L. parisiensis}, \textit{L. israelensis}, \textit{L. gormanii}, \textit{L. longbeachae}, and of \textit{Escherichia coli}. All of the measured species possess a positive acoustic contrast, leading to the focusing into a pressure node of a standing wave. A broad variation in the shape and size across the species, as well as within each species, considerably affects the magnitude of the acoustic radiation force, resulting in observable differences in the rate of focusing among the species.
Since \textit{Legionella} are known to infect protozoa, such as \textit{Acanthamoeba} species,\cite{bartram2007legionella} and grow within, we also measure the acoustic contrast of \textit{A. castellanii} in cyst and trophozoite states, co-cultured with \textit{L. pneumophila}. \textit{A. castellanii} generally also feature a positive acoustic contrast factor and thus focus into a pressure node. Encysted \textit{A. castellanii} exhibit a significantly higher acoustic contrast factor than the trophozoites.

To complement experimental observations, we introduce a multi-body dynamics algorithm for investigating the role of the experimentally-observed cell size distribution and cell concentration in the collective acoustophoresis of bacteria. The algorithm is based on Stokesian dynamics for modeling the hydrodynamic interactions,\cite{durlofsky1987dynamic} supplemented by the Brownian motion,\cite{brady1993brownian} acoustic radiation forces\cite{yosioka1955acoustic} and acoustic interaction forces.\cite{silva2014acoustic} The simulations reveal that an increase in the cell concentration accelerates the rate of acoustic focusing due to hydrodynamic interactions. This phenomenon, which we call \textit{collective hydrodynamic focusing (CHF)}, does not significantly influence our measurements, but could improve the understanding of many other studies and applications that feature an externally-forced motion of bacteria-sized objects, such as acoustic seed-particle trapping of bacteria and nanoparticles,\cite{hammarstrom2012seed} bacteria separation using ultrasound,\cite{miles1995principles} or capturing of paramagnetic beads using magnetic field gradient.\cite{mikkelsen2005microfluidic}

\section*{Results}

\begin{figure*}[h]
\centering
\includegraphics[scale=\scaleFactor]{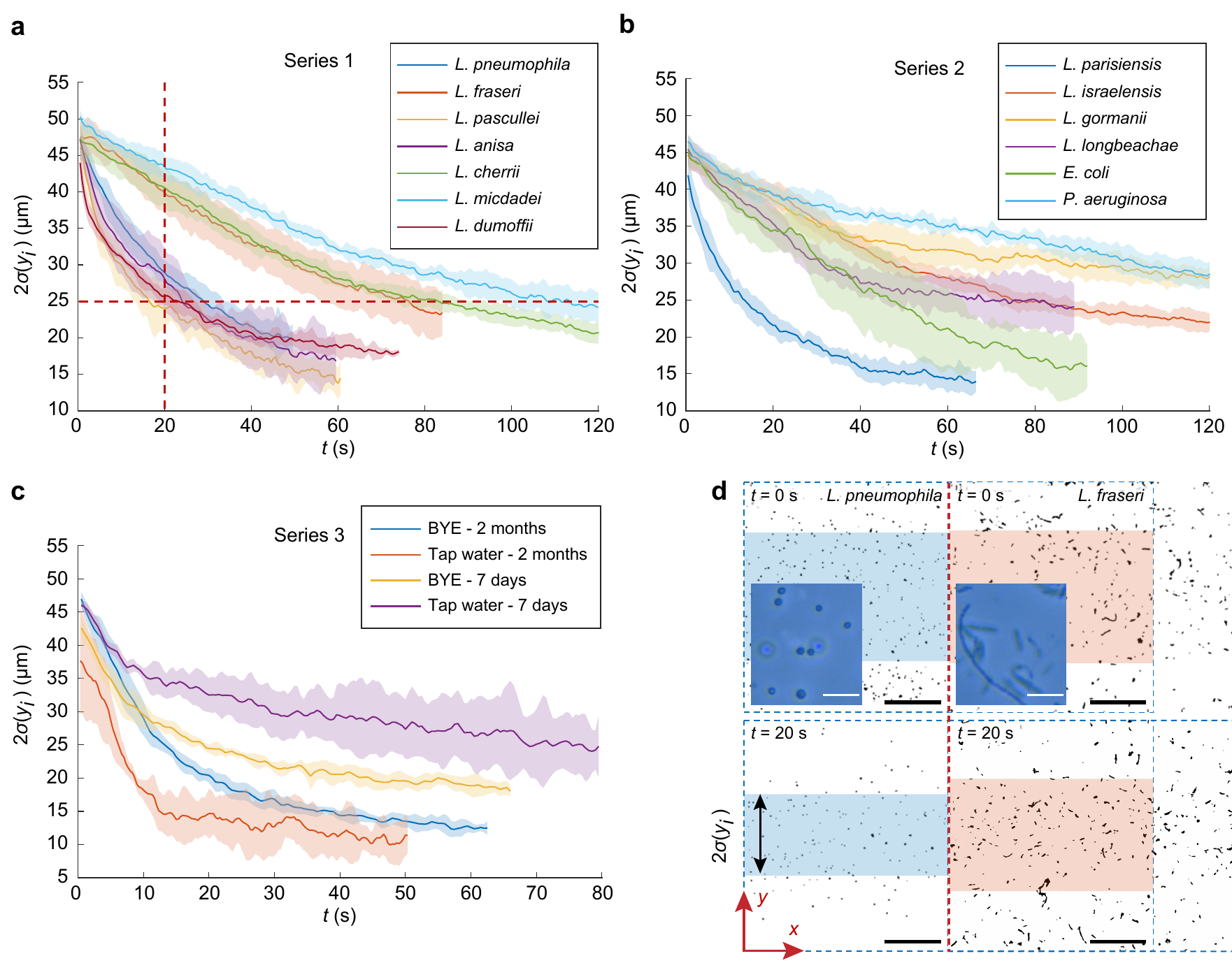}
\caption{Acoustic focusing of \textit{Legionella} species, \textit{E. coli}, and \textit{P. aeruginosa} in a standing acoustic wave. (a) \& (b) Time evolution of line widths of all the investigated species, from two series of focusing experiments; $p_{\mathrm{a}} = 210 \pm \SI{8.4}{\kilo\pascal}$ in series 1 and $p_{\mathrm{a}} = 190 \pm \SI{13.7}{\kilo\pascal}$ in series 2, while the frequency was constant across the two series at $f=\SI{3829}{\kilo\hertz}$. (c) Time evolution of line widths of \textit{L. pneumophila} subjected to different conditioning protocols; $p_{\mathrm{a}} = 352 \pm \SI{19.7}{\kilo\pascal}$ and $f=\SI{3941}{\kilo\hertz}$. (d) Snapshots of segmented binary images of \textit{L. pneumophila} and \textit{L. fraseri} at $t=\SI{0}{\second}$ and $t=\SI{20}{\second}$, and comparison of the line widths (shaded areas) between the two species, defined as two standard deviations $\sigma \left( y_i \right)$ of the $y$-position of all the bacteria in the observed frame of $\SI{292}{\micro\meter} \times \SI{164}{\micro\meter}$ (the snapshots are half of the actually analyzed length in the $x$-direction). The insets show a large difference in the morphology of the two subspecies; white scale bar corresponds to $\SI{5}{\micro\meter}$, black scale bar corresponds to $\SI{40}{\micro\meter}$. The line width curves in (a)-(c) represent a mean of $2 \sigma \left( y_i \right)$ from 4-8 repetitions of experiments; shaded area surrounding the curves is a standard deviation of the mean $2 \sigma \left( y_i \right)$ across the repetitions. The vertical dashed line in (a) represents a reference timestamp at $t=\SI{20}{\second}$, while the horizontal dashed line corresponds to a reference line width of $2 \sigma \left( y_i \right)= \SI{25}{\micro\meter}$.}
\label{fig:time_lapse_exp_results}
\end{figure*}

\begin{figure*}
\centering
\includegraphics[scale=\scaleFactor]{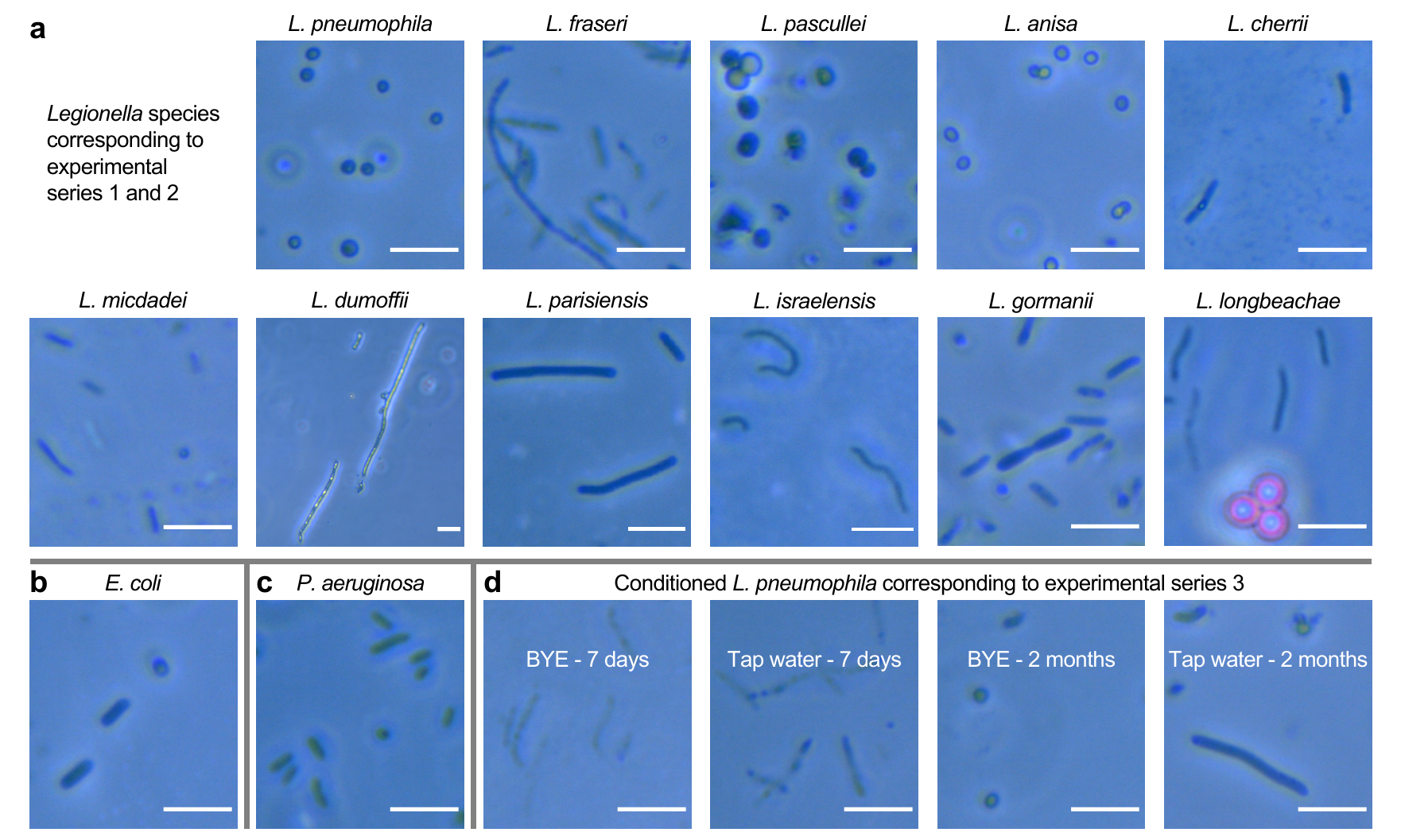}
\caption{Experimentally observed variation in shape and size between investigated bacterial species. (a) \textit{Legionella} spp. investigated in experimental series 1 and 2; (b) \textit{E. coli} and (c) \textit{P. aeruginosa} investigated in series 2; (d) conditioned \textit{L. pneumophila} investigated in series 3. Images are obtained by a high-magnification (1000x) phase-contrast microscopy. Red spheres next to \textit{L. longbeachae} are reference polystyrene particles with the diameter of $\SI{2.21}{\micro\meter}$. Scale bar corresponds to $\SI{5}{\micro\meter}$.}
\label{fig:phi_size_exp_results}
\end{figure*}

\subsubsection*{Definition of an effective acoustic contrast factor}

%%% NEW SECTION ON THE EFFECTIVE AC. CONTRAST %%%
During acoustophoresis, the motion of an object of mass $m$ is governed by Newton's second law of motion
\begin{equation}
  m \frac{\mathrm{d} \boldsymbol{v}_{\mathrm{obj}}}{\mathrm{d} t} = \boldsymbol{F}_{\mathrm{S}} + \boldsymbol{F}_\mathrm{ARF},
  \label{eq:motion}
\end{equation}
with the velocity of the object $\boldsymbol{v}_{\mathrm{obj}}$, the Stokes drag $\boldsymbol{F}_{\mathrm{S}}$, and the ARF $\boldsymbol{F}_{\mathrm{ARF}}$.
For length scales applicable in our work, it is sufficient to approximate the motion of such an object by assuming it reaches its terminal velocity instantaneously,\cite{bruus2012arf} resulting in the force equilibrium
\begin{equation}
  \boldsymbol{F}_\mathrm{ARF} = - \boldsymbol{F}_{\mathrm{S}}.
  \label{eq:equilibriumForce}
\end{equation}

The ARF on an object featuring a plane of symmetry in a plane standing wave in an inviscid fluid that is small in size compared to the acoustic wavelength $\lambda = c_0/f$, with the speed of sound in the fluid $c_0$, can be generally written in the following form,\cite{sepehrirahnama2021acoustic}
\begin{equation}
    \boldsymbol{F}_{\mathrm{ARF}} = 3 V \Phi k E_{\mathrm{ac}} \sin \left( 2 k y \right) \boldsymbol{e}_y , \label{eq:arf}
\end{equation}
with the object's volume $V$, the wavenumber $k$, the position of the object in the wave $y$ along the direction of the unit vector $\boldsymbol{e}_y$ that is parallel to the pressure gradient of the wave and normal to a plane of symmetry of the object, and the acoustic energy density $E_{\mathrm{ac}} = p_{\mathrm{a}}^2/\left( 4 \rho_0 c_0^2 \right)$, with the equilibrium fluid density $\rho_0$ and the pressure amplitude $p_{\mathrm{a}}$. The acoustic contrast factor $\Phi$ represents the acoustic scattering at the object and is consequently shape-, orientation- and material-dependent. For a homogeneous sphere, $\Phi$ can be expressed as\cite{yosioka1955acoustic,bruus2012arf}
\begin{equation}
    \Phi^{\mathrm{sph}} = \frac{1}{3} \left[ \frac{5 \rho_{\mathrm{obj}} - 2 \rho_0}{2 \rho_{\mathrm{obj}} + \rho_0} - \frac{\kappa_{\mathrm{obj}}}{\kappa_0} \right]
\end{equation}
with the density of the object $\rho_{\mathrm{obj}}$, and the compressibility of the particle and the fluid, $\kappa_{\mathrm{obj}}$ and $\kappa_0$, respectively. For a long cylinder with its axis parallel to a pressure nodal plane, $\Phi$ follows as\cite{wei2004acoustic}
\begin{equation}
    \Phi^{\mathrm{cyl}} = \frac{1}{3} \left[ \frac{4 \rho_{\mathrm{obj}}}{2 \rho_{\mathrm{obj}} + 2 \rho_0} - \frac{\kappa_{\mathrm{obj}}}{\kappa_0} \right] .
\end{equation}
Considering a polystyrene\cite{selfridge1985approximate} object in water results in $\Phi^{\mathrm{sph}} = 0.1707$ and $\Phi^{\mathrm{cyl}} = 0.1709$; the difference between $\Phi^{\mathrm{sph}}$ and $\Phi^{\mathrm{cyl}}$ grows with the density contrast, but remains negligible in the density range associated with biological cells. With the density contrast, viscous contributions to the ARF also rise and can be absorbed into $\Phi$, since only the overall ARF scaling is affected while the spatial dependency with $y$ remains unchanged.\cite{pavlic2022shapes} For objects of various shapes, sizes, and materials, $\Phi$ that accounts for the viscosity can be extracted from a recent numerical investigation.\cite{pavlic2022shapes}

In addition to the ARF, a moving object in an acoustic field is subjected to the Stokes drag\cite{happel2012low} due to the acoustic streaming velocity $\left< \boldsymbol{v}_2 \right>$ and the velocity of the object itself, namely,
\begin{equation}
  \boldsymbol{F}_{\mathrm{S}} = 6 \pi \eta \mathrm{ESR} \left[ \boldsymbol{K} \boldsymbol{\cdot} \left( \left< \boldsymbol{v}_2 \right> - \boldsymbol{v}_{\mathrm{obj}} \right) \right],
  \label{eq:Fstream}
\end{equation}
with the shape-, orientation, and configuration-dependent translation tensor $\boldsymbol{K}$ that depends also on the surrounding objects and walls, the equivalent sphere radius $\mathrm{ESR}$, while neglecting the contribution from rotations. $\boldsymbol{K}$ is implicitly normalized by $6 \pi \mathrm{ESR}$, compared to the dimensional $\boldsymbol{K}$ in Happel and Brenner.\cite{happel2012low} Since we are interested only in the force component along the $y$-axis and since we did not observe significant motion in other directions ($\boldsymbol{v}_{\mathrm{obj}} \boldsymbol{\cdot} \boldsymbol{e}_x \approx \boldsymbol{v}_{\mathrm{obj}} \boldsymbol{\cdot} \boldsymbol{e}_z \approx 0$), neglecting the streaming velocity ($\left< \boldsymbol{v}_2 \right> \approx \boldsymbol{0}$), we can write
\begin{equation}
    \boldsymbol{F}_{\mathrm{S}} \boldsymbol{\cdot} \boldsymbol{e}_y = - 6 \pi \eta \mathrm{ESR} K_{yy} \left( \boldsymbol{v}_{\mathrm{obj}} \boldsymbol{\cdot} \boldsymbol{e}_y \right) , \label{eq:FdragKyy}
\end{equation}
where $K_{yy}$ can be interpreted as a correction to the Stokes drag on a sphere due to the shape, orientation, and configuration of the system (i.e. $K_{yy}=1$ for an isolated sphere in an unbounded medium). The influence of walls on $K_{yy}$ of $\SI{}{\micro\meter}$-sized particles in a typical acoustofluidic device was quantified by \citet{barnkob2012acoustic}, while \citet{happel2012low} quantified the influence of the nearby objects in a few representative configurations.

For a plane standing wave, the velocity of an object undergoing acoustophoresis can be expressed as
\begin{equation}
    \boldsymbol{v}_{\mathrm{obj}} \boldsymbol{\cdot} \boldsymbol{e}_y = \frac{\mathrm{ESR}^2 \Phi_{\mathrm{eff}} k p_{\mathrm{a}}^2}{6 \eta \rho_0 c_0^2} \sin \left( 2 k y \right) ,
    \label{legio:eq:velCell}
\end{equation}
with the effective acoustic contrast factor $\Phi_{\mathrm{eff}}$ defined as
\begin{equation}
  \Phi_{\mathrm{eff}} = \frac{\Phi}{K_{yy}},
  \label{legio:eq:PhiEffDef}
\end{equation}
which depends on the material properties, orientation and shape of the object, configuration of the system, as well as viscous contributions to the ARF. As such, $\Phi_{\mathrm{eff}}$ is not necessarily constant in $y$, since distances to walls and other objects vary with the motion of the object.
%%%%%%%%%%%%%%%%%%%%%%%%%%%%%%%%%%%%%%%%%%%%%%%%%

In the experimental setting, we use a glass-silicon-glass bulk-acoustic-wave (BAW) device shown in Fig. \ref{fig:device_setup}(a)-(c), driven by a piezoelectric transducer at $\SI{}{\mega\hertz}$-frequencies, to generate a one-dimensional standing acoustic wave across the width of the water-filled channel. The see-through channel design is chosen to avoid the need for any kind of staining the bacteria, since the influence of staining on the acoustic contrast factor is unknown. The standing wave is experimentally characterized through $p_{\mathrm{a}}$ and $k$, which are, as indicated in Fig. \ref{fig:device_setup}(d), measured by tracking the motion of $\SI{4.97}{\micro\meter}$ polystyrene (PS) particles with known material properties, and fitting the theoretical trajectory of a particle\cite{barnkob2010measuring,barnkob2012acoustic,bruus2012arf}
\begin{equation}
    y_i (t) = \frac{1}{k} \arctan \left\{ \tan \left[ k y_i (t = 0) \right] \exp \left[ \frac{\Phi_{\mathrm{eff}} p_{\mathrm{a}}^2}{3 \rhofl \cfl^2 \eta} k \mathrm{ESR}^2 t \right] \right\} 
    \label{eq:theoretTraject}
\end{equation}
that is found by integrating eq. (\ref{legio:eq:velCell}) in time, to the experimental trajectory, using $p_{\mathrm{a}}$ and $k$ as the fitting parameters and assuming $\Phi_{\mathrm{eff}} = \Phi^{\mathrm{sph}}$.
The effective acoustic contrast factor of a cell is measured by tracking the motion of individual cells in the characterized standing wave and then fitting the theoretical trajectory from eq. (\ref{eq:theoretTraject}) to the experimental trajectory, using $\Phi_{\mathrm{eff}}$ as the fitting parameter. The $\mathrm{ESR}$ is for amoebae specific to an individual cell, while for bacteria, the average ESR of an individual species is used for all $\Phi_{\mathrm{eff}}$ measurements within the species.
The spatial dependence of $\Phi_{\mathrm{eff}}$ is investigated in the supplemental material by measuring $\boldsymbol{v}_{\mathrm{obj}}$ of bacteria at various $y$-positions, and inserting the measured velocities into eq. (\ref{legio:eq:velCell}).

\subsection*{Acoustophoresis of bacteria}
Transient behaviour of cells during acoustic focusing is shown in Fig. \ref{fig:time_lapse_exp_results} for all bacterial species that were investigated. The behaviour is quantified through the line width $2 \sigma \left( y_{\mathrm{i}} \right)$, where $\sigma \left( y_{\mathrm{i}} \right)$ is the standard deviation of the $y$-positions of all the bacteria that are within the $\SI{292}{\micro\meter} \times \SI{164}{\micro\meter}$ frame under observation. The line width is for each species averaged across 4-10 repetitions of the experiment. The results are split into two series, shown in Fig. \ref{fig:time_lapse_exp_results}(a) and \ref{fig:time_lapse_exp_results}(b), corresponding to two separate experimental sessions, at $f=\SI{3829}{\kilo\hertz}$ and with the average pressure amplitudes of $210 \pm \SI{8.4}{\kilo\pascal}$ in series 1 and $190 \pm \SI{13.7}{\kilo\pascal}$ in series 2. In the experiments of series 3 ($f=\SI{3941}{\kilo\hertz}$, $p_{\mathrm{a}} = 352 \pm \SI{19.7}{\kilo\pascal}$) in Fig. \ref{fig:time_lapse_exp_results}(c), the acoustic focusing is analyzed for samples of \textit{L. pneumophila} that were subjected to different conditioning protocols.

Figures \ref{fig:time_lapse_exp_results}(a) and \ref{fig:time_lapse_exp_results}(b) indicate that across the tested \textit{Legionella} species, rate of focusing varies significantly. Surprisingly, even subspecies \textit{L. pneumophila}, \textit{L. fraseri}, and \textit{L. pascullei} considerably differ in their acoustic focusability, reaching for example $2\sigma \left( y_{\mathrm{i}} \right) = \SI{25}{\micro\meter}$ at $\SI{29}{\second}$, $\SI{76}{\second}$, and $\SI{16.5}{\second}$, respectively. The large spread in the observed behaviour could be due to variation in the volume, shape, material properties, or cell concentration. Significant differences in cell morphology across the species and subspecies, observable in Fig. \ref{fig:phi_size_exp_results}, indicate potentially large influence of the morphology on the rate of focusing. However, an important observation is that all the \textit{Legionella} species, as well as \textit{E. coli} and \textit{P. aeruginosa}, experience observable acoustic radiation force that drives the focusing of cells towards the pressure node, despite the large variability in the rate of focusing.

\begin{figure*}
\centering
\includegraphics[scale=\scaleFactor]{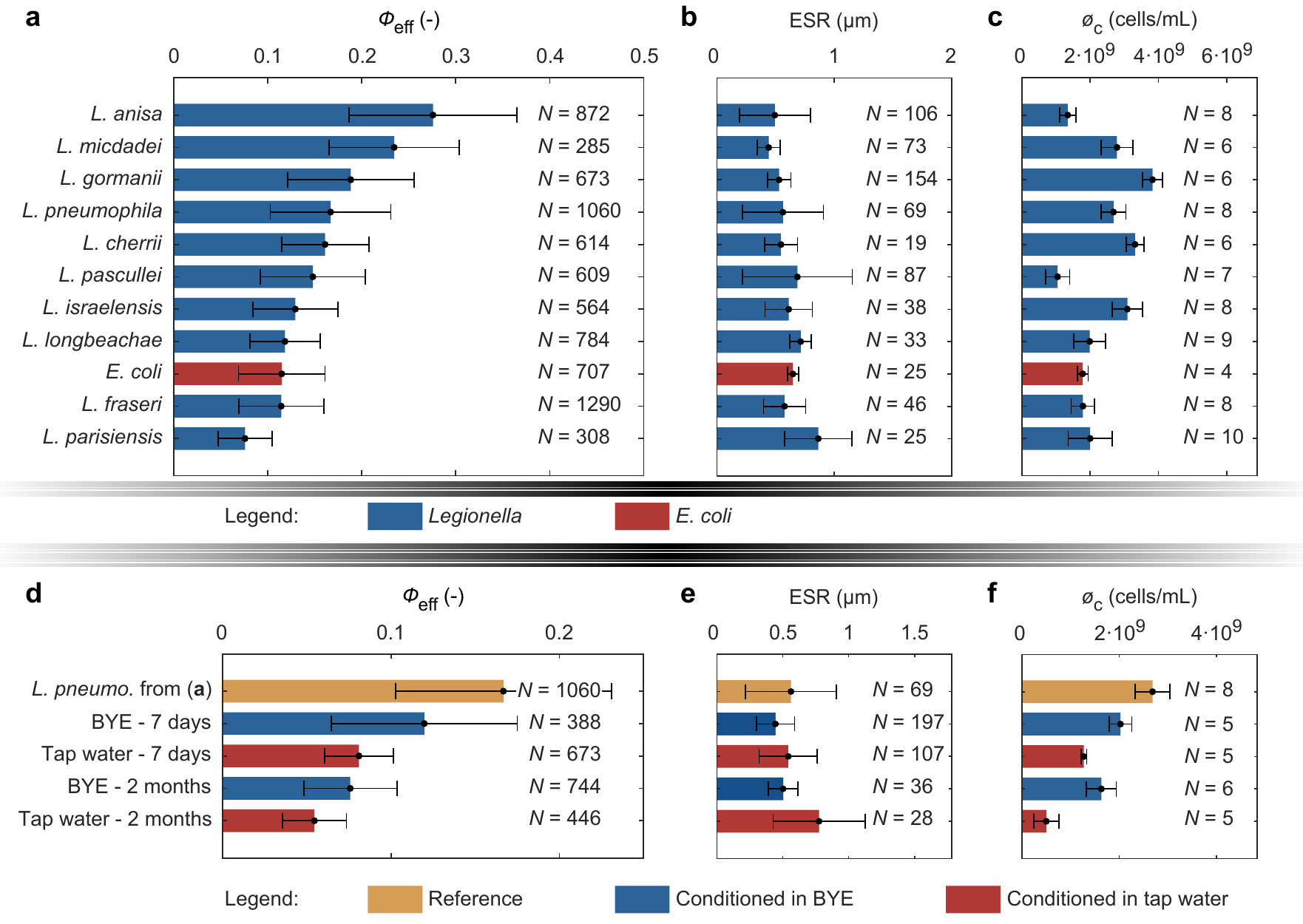}
\caption{Experimentally determined effective acoustic contrast factor $\Phi_{\mathrm{eff}}$ of the bacterial species. (a) Mean $\Phi_{\mathrm{eff}}$ for all the bacterial species, averaged over $N = 285 - 1290$ cell trajectories, obtained from 4-8 experiments per species. (b) The corresponding equivalent sphere radius (ESR) measured from high-magnification images (1000x) for $N = 19 - 154$ cells per species. (c) The corresponding initial cell concentration $\phi_{\mathrm{c}}$. (d) Mean $\Phi_{\mathrm{eff}}$ for \textit{L. pneumophila} conditioned in BYE and tap water, for 7 days and for 2 months. (e) The corresponding ESR measured from high-magnification images (1000x) for $N = 28 - 197$ cells per sample. (f) The corresponding $\phi_{\mathrm{c}}$. Errorbars in the figure represent the standard deviation from the mean value.}
\label{fig:phi_exp_results}
\end{figure*}

\begin{figure*}
\centering
\includegraphics[scale=\scaleFactor]{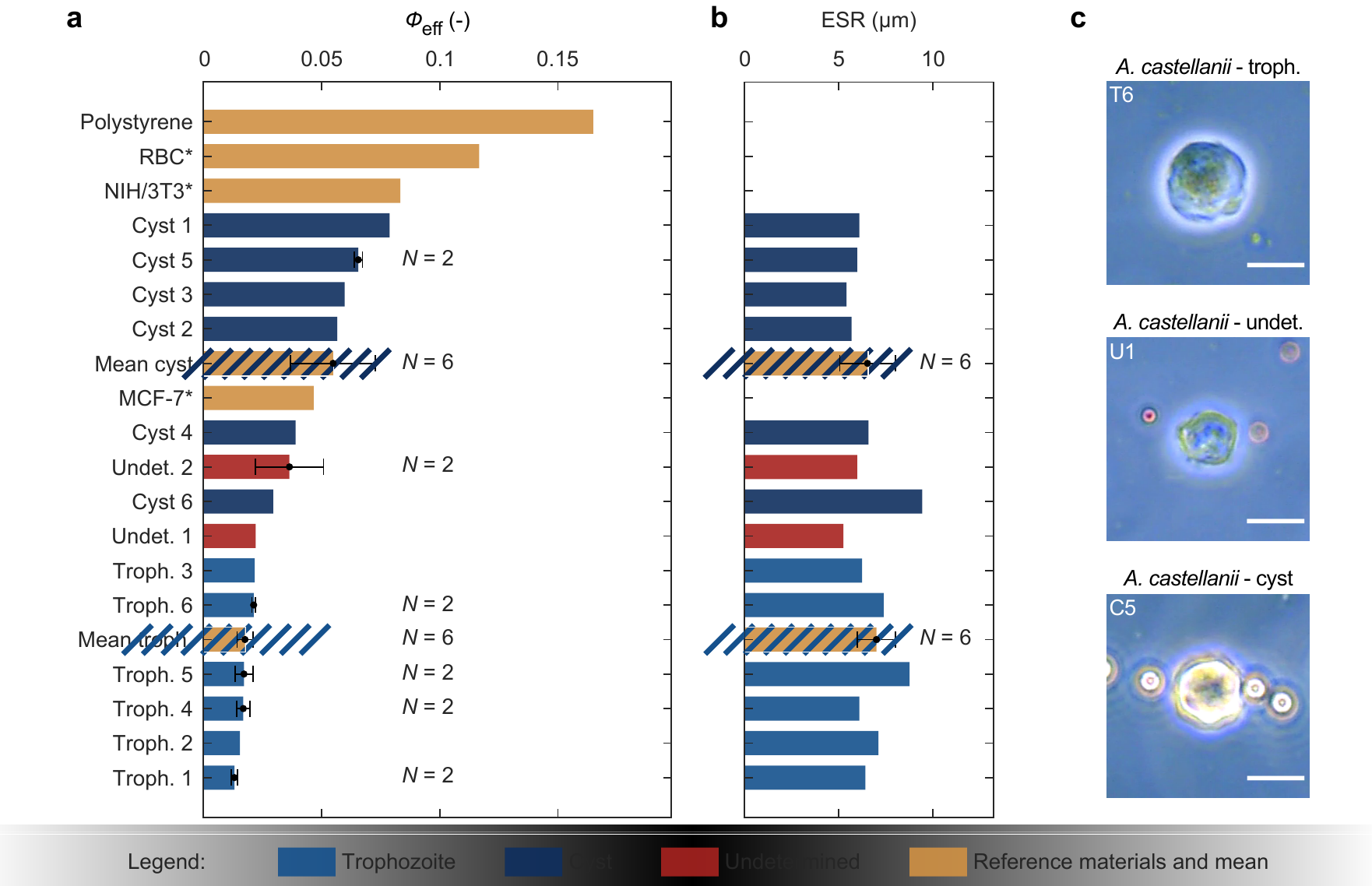}
\caption{Experimentally determined effective acoustic contrast factor $\Phi_{\mathrm{eff}}$ of \textit{A. castellanii} co-cultured with \textit{L. pneumophila}. (a) Mean $\Phi_{\mathrm{eff}}$ for the individual amoebae, averaged over $N = 1 - 2$ cell trajectories ($N = 6$ for the mean values). Amoebae were subjected to a standing wave with $f=\SI{1987}{\kilo\hertz}$. Reference single-cell acoustic contrast factors denoted with $^{*}$ are obtained from \citet{hartono2011chip}, while the contrast of polystyrene is based on the reported material properties.\cite{selfridge1985approximate} (b) The corresponding equivalent sphere radius (ESR). (c) Morphology of the three visually distinguished states of \textit{A. castellanii}, namely, trophozoite, cyst, and an intermediate state (undetermined). Red and white spheres next to amoebae are reference polystyrene particles (diameter of $\SI{2.21}{\micro\meter}$ and $\SI{4.97}{\micro\meter}$, respectively). Scale bar corresponds to $\SI{10}{\micro\meter}$.}
\label{fig:phi_amoeba_exp_results}
\end{figure*}

The trajectories of individual bacteria from experiments in Fig. \ref{fig:time_lapse_exp_results}(a)-(b) were used to measure the effective acoustic contrast factor $\Phi_{\mathrm{eff}}$ of the bacterial species, shown in Fig. \ref{fig:phi_exp_results}(a). The mean measured $\Phi_{\mathrm{eff}}$ varies between $0.28 \pm 0.089$ for \textit{L. anisa} and $0.076 \pm 0.029$ for \textit{L. parisiensis}. In terms of the rate of focusing from Fig. \ref{fig:time_lapse_exp_results}, \textit{L. parisiensis} focuses significantly faster than \textit{L. micdadei} ($\Phi_{\mathrm{eff}} = 0.23 \pm 0.069$), despite the greater $\Phi_{\mathrm{eff}}$ of the latter and the lower $p_{\mathrm{a}}$ in experimental series 2. The faster focusing of \textit{L. parisiensis} can be attributed to the larger volume ($\mathrm{ESR} = 0.86 \pm \SI{0.29}{\micro\meter}$) compared to \textit{L. micdadei} ($\mathrm{ESR} = 0.44 \pm \SI{0.097}{\micro\meter}$), since $\boldsymbol{v}_{\mathrm{obj}} \propto \Phi_{\mathrm{eff}} \mathrm{ESR}^2$.
\textit{P. aeruginosa} and \textit{L. dumoffii} were excluded from the $\Phi_{\mathrm{eff}}$-analysis, due to excessive swimming motion of the former and the significantly elongated shape of the latter, as depicted in Fig. \ref{fig:phi_size_exp_results}.
Size-wise, the mean ESR across the bacterial species varies between $0.44 \pm \SI{0.097}{\micro\meter}$ for \textit{L. micdadei} and $0.86 \pm \SI{0.29}{\micro\meter}$ for \textit{L. parisiensis}, and generally decreases with an increase in the measured $\Phi_{\mathrm{eff}}$.
%On average, species that feature larger $\Phi_{\mathrm{effM}}$, also feature higher cell concentrations shown in Fig. \ref{fig:phi_exp_results}(c). However, there is no clear indication from the literature that higher concentration could influence the acoustic focusability.

In the environment (e.g. in water installations), \textit{Legionella} can enter a so-called viable but non-culturable (VBNC) state, in which they are still dangerous for humans, while also very difficult to detect using standard methods.\cite{dietersdorfer2018starved} In Fig. \ref{fig:time_lapse_exp_results}(c) and Fig. \ref{fig:phi_exp_results}(d)-(f), we compare the acoustophoresis among the same strain of \textit{L. pneumophila}, but exposed to various conditions that trigger VBNC state through starvation: cells conditioned in sterilized tap water, compared to cells conditioned in buffered yeast extract (BYE) and transferred into the sterilized tap water shortly before the experiment. Measurements show that cells conditioned in tap water exhibit larger mean ESR %and lower mean $\phi_{\mathrm{c}}$
than the cells conditioned in BYE. Analogously to the general relation of $\Phi_{\mathrm{eff}}$ from Fig. \ref{fig:phi_exp_results}(a)-(b), mean $\Phi_{\mathrm{eff}}$ decreases with the increase in the mean ESR. %and decreases with the decrease in the cell concentration.
Prolonged conditioning (2 months compared to 7 days) enhances this trend. Overall, the conditioning influences the magnitude of $\Phi_{\mathrm{eff}}$, but the cells are in all cases focusing towards the pressure node, as indicated in Fig. \ref{fig:time_lapse_exp_results}(c).

\subsection*{Acoustophoresis of amoebae}
Measurements of the effective acoustic contrast factor $\Phi_{\mathrm{eff}}$ for 14 amoebae are presented in Fig. \ref{fig:phi_amoeba_exp_results}. The amoebae were classified into trophozoites, cysts, and undetermined (amoeba in transition from a trophozoite to a cyst), based on a visual observation of the presence of the cellulose wall that is characteristic for cysts.\cite{bowers1969fine} The results indicate that the cysts have a higher acoustic contrast factor ($\Phi_{\mathrm{eff}} = 0.055 \pm 0.018$) than the trophozoites ($\Phi_{\mathrm{eff}} = 0.018 \pm 0.003$), while the undetermined are in-between ($\Phi_{\mathrm{eff}} = 0.029 \pm 0.010$). This is reasonable also from a theoretical perspective, since the wall of a cyst contains cellulose which is denser ($\rho \approx \SI{1.6}{\gram\per\cubic\centi\meter}$)\cite{sun2005true} and less compressible ($\kappa \approx 5\times\SI[parse-numbers=false]{10^{-11}}{\per\pascal}$)\cite{quesada2011nanomechanical} than a cell-like material ($\rho \approx \SI{1.1}{\gram\per\cubic\centi\meter}$ and $\kappa \approx 3.3\times\SI[parse-numbers=false]{10^{-10}}{\per\pascal}$ for a red blood cell)\cite{hartono2011chip}, leading to an increase in the effective density and to a decrease in the effective compressibility of a cell, increasing its $\Phi_{\mathrm{eff}}$.

\begin{figure}[h]
\centering
\includegraphics[scale=\scaleFactor]{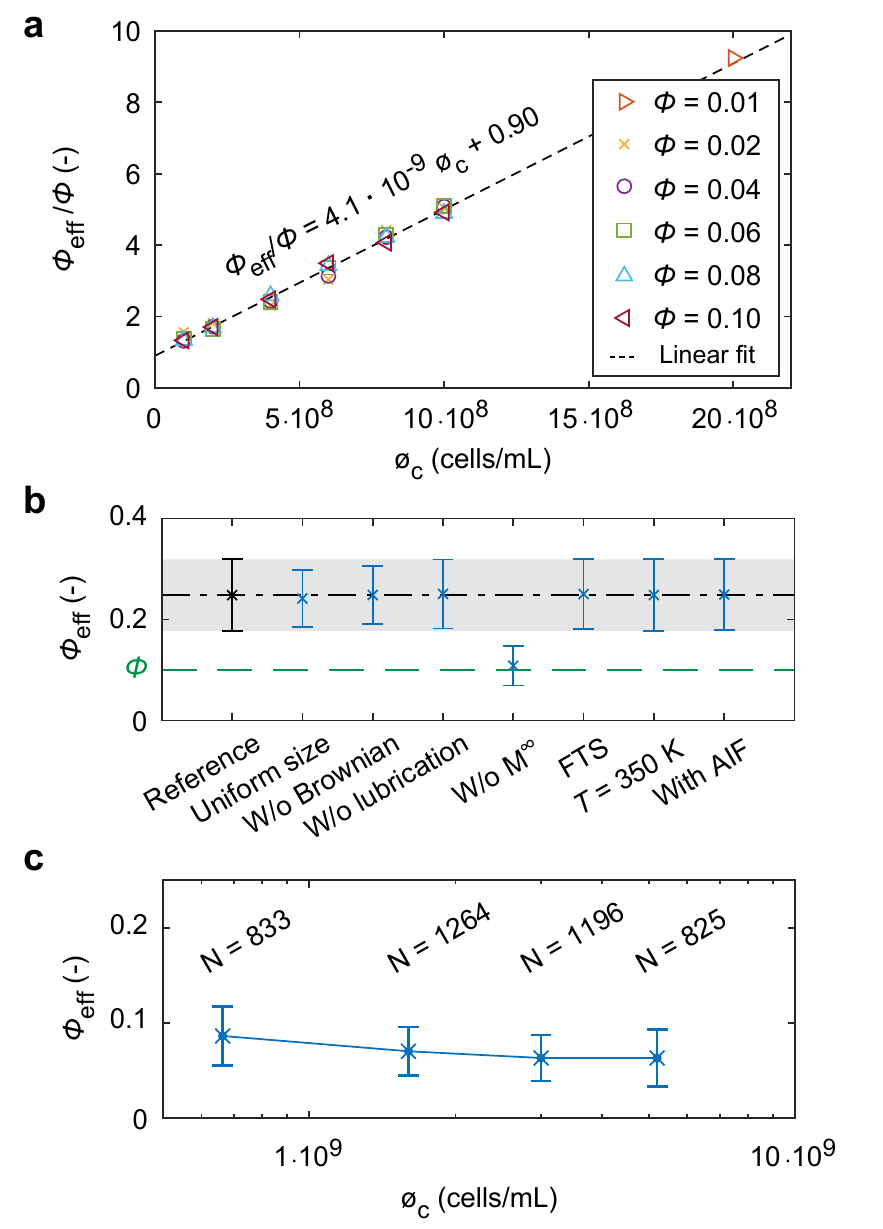}
\caption{Influence of collective hydrodynamic focusing (CHF) in computations and experiments. (a) Extracted ratio between the multi-body $\Phi_{\mathrm{eff}}$ and an assigned single-body $\Phi$ from the multi-body numerical simulations in dependence of the initial cell concentration $\phi_{\mathrm{c}}$, for the experimental particle size distribution of \textit{L. pneumophila}. (b) Influence of individual contributions to the computational model on the extracted $\Phi_{\mathrm{eff}}$ for $\Phi=0.1$ (dashed line) and $\phi_{\mathrm{c}}=4\cdot 10^8 \mathrm{cells}/\SI{}{\milli\liter}$ (reference - the mean $\Phi_{\mathrm{eff}}$ is given by the dash-dotted line): assigning the mean ESR of \textit{L. pneumophila} to all the particles in the simulation (uniform size); neglecting the Brownian motion (w/o Brownian); neglecting the near-field lubrication interactions (w/o lubrication); neglecting the many-body hydrodynamic interactions (w/o $\mathrm{M}^{\infty}$); using a more accurate FTS algorithm\cite{durlofsky1987dynamic} of Stokesian dynamics that also accounts for stresslets (FTS); increasing the temperature that increases the Brownian motion ($T=\SI{350}{\kelvin}$); including the inviscid acoustic interactions\cite{silva2014acoustic} (with AIF). (c) Experimental estimation of the influence of $\phi_{\mathrm{c}}$ on the measured $\Phi_{\mathrm{eff}}$ for a series of experiments with \textit{L. pneumophila} in a $\SI{200}{\micro\meter}$ wide channel, performed at $f=\SI{3930.9}{\kilo\hertz}$, with $p_{\mathrm{a}} = 382 \pm \SI{30.1}{\kilo\pascal}$ and $\mathrm{ESR} = 0.69 \pm \SI{0.17}{\micro\meter}$.}
\label{fig:stokesian_results}
\end{figure}

In our experiments, the cysts with $\mathrm{ESR} = 6.5 \pm \SI{1.5}{\micro\meter}$) appear marginally smaller than the trophozoites with $\mathrm{ESR}=7.0 \pm \SI{1.0}{\micro\meter}$. Since the trajectory of an amoeba in the standing acoustic wave depends on the factor $\Phi_{\mathrm{eff}} \mathrm{ESR}^2$, which amounts to $\SI{0.86}{\micro\meter\squared}$ for trophozoites and $\SI{2.34}{\micro\meter\squared}$ for cysts, the rate of focusing of cysts is significantly higher than that of trophozoites.

\subsection*{Multi-body effects}
The comparison across species of \textit{Legionella} from Fig. \ref{fig:phi_exp_results} indicates that the initial concentration of cells in the sample $\phi_{\mathrm{c}}$ could influence the measured $\Phi_{\mathrm{eff}}$. %since $\Phi_{\mathrm{effM}} = 0.396$ of \textit{L. micdadei} exceeds a theoretical upper bound of $\Phi = 0.38$ that follows from eq. (\ref{eq:Phi}) by assuming the rigid particle limit $\kappa_{\mathrm{p}} \ll \kappa_0$ and $\SI{1.156}{\gram\per\cubic\centi\meter}$ for the density of bacteria.
However, in the existing studies of acoustophoresis of bacteria or bacteria-sized particles,\cite{antfolk2014focusing,gerlt2022focusing,devendran2020diffraction,ugawa2022reduced,hawkes1997filtration} the multi-body effects on the acoustic focusing are of unknown magnitude and generally neglected, despite the concentrations of cells or particles in such experiments often surpassing $10^8 \mathrm{cells}/\SI{}{\milli\liter}$.

In Fig. \ref{fig:stokesian_results}, we investigate the influence of the cell concentration of \textit{L. pneumophila} on the measured acoustic contrast $\Phi_{\mathrm{eff}}$ through multi-body computations, under conditions corresponding to our experiments ($f=\SI{3829}{\kilo\hertz}$, $p_{\mathrm{a}} = \SI{210}{\kilo\pascal}$, channel width of $\SI{200}{\micro\meter}$). The multi-body simulations for $10^{8} \mathrm{cells}/\SI{}{\milli\liter}\leq \phi_{\mathrm{c}} \leq 10^{9} \mathrm{cells}/\SI{}{\milli\liter}$ and for the assigned single-cell acoustic contrast factor $\Phi$ between $0.01$ and $0.1$, reveal a significant dependence of the measured $\Phi_{\mathrm{eff}}$ on $\phi_{\mathrm{c}}$ in Fig. \ref{fig:stokesian_results}(a). We analyze this relation through the ratio $\Phi_{\mathrm{eff}}/\Phi$ that quantifies the multi-body effects ($\Phi_{\mathrm{eff}}/\Phi=1$ means a negligible influence of multi-body effects), which shows a linear dependence on the cell concentration in $\mathrm{cells}/\SI{}{\milli\liter}$: $\Phi_{\mathrm{eff}}/\Phi = 4.1 \cdot 10^{-9} \phi_{\mathrm{c}} + 0.90$. A higher concentration simulation, at $\phi_{\mathrm{c}}=2\cdot10^9 \mathrm{cells}/\SI{}{\milli\liter}$, fits on the linear relation relatively well.

To determine the mechanism responsible for the amplification of $\Phi_{\mathrm{eff}}$ as $\phi_{\mathrm{c}}$ is increased, we turn off or modify individual contributions in our computational model. The results shown in Fig. \ref{fig:stokesian_results}(b) indicate that the influences of the experimentally-determined cell size distribution, Brownian motion, near-field lubrication interactions, and the acoustic interaction force (AIF) are negligible compared to the many-body hydrodynamic interactions ($\mathrm{M}^{\infty}$). Turning the latter off results in $\Phi_{\mathrm{eff}}$ that is very close to the single-particle $\Phi$ of $0.1$; furthermore, turning off other contributions to the dynamics of bacteria has negligible influence on the extracted $\Phi_{\mathrm{eff}}$.

To relate the simulations to our $\Phi_{\mathrm{eff}}$ measurements, we perform a series of experiments with our standard experimental setup for bacteria that was used in Fig. \ref{fig:phi_exp_results}. The results for \textit{L. pneumophila} with $6.6\cdot10^8 \mathrm{cells}/\SI{}{\milli\liter} \leq \phi_{\mathrm{c}} \leq 5.2\cdot10^9 \mathrm{cells}/\SI{}{\milli\liter}$, presented in Fig. \ref{fig:stokesian_results}(c), indicate that cell-concentration-related multi-body effects are negligible in our experiments.

\subsection*{Swimmers in acoustophoresis}
To contain a bacterium with a given maximal swimming speed of $v_{\mathrm{bac}}^{\mathrm{max}}$ within the one-dimensional acoustic trap, the pressure amplitude has to exceed a certain threshold ($p_{\mathrm{a}}^{\mathrm{min}}$). $v_{\mathrm{bac}}^{\mathrm{max}}$ defines a swimming thrust through the balance with the Stokes drag on the bacterium, and balancing the swimming thrust with the ARF gives the approximation for $p_{\mathrm{a}}^{\mathrm{min}}$, namely,
\begin{equation}
    p_{\mathrm{a}}^{\mathrm{min}} = \sqrt{\frac{3 \eta \rho_0 c_0^3}{\pi f \mathrm{ESR}^2 \Phi_{\mathrm{eff}}} v_{\mathrm{bac}}^{\mathrm{max}}} ,
\end{equation}
resulting in $p_{\mathrm{a}}^{\mathrm{min}} \approx \SI{667}{\kilo\pascal}$ for \textit{E. coli} ($\Phi_{\mathrm{eff}} = 0.11 \pm 0.046$, $\mathrm{ESR} = 0.65 \pm \SI{0.047}{\micro\meter}$) with $v_{\mathrm{bac}}^{\mathrm{max}} = \SI{25}{\micro\meter\per\second}$\cite{wadhwa2021bacterial} and $f = \SI{3.9}{\mega\hertz}$, assuming a spherical shape.

In the acoustophoresis experiments, we observed swimming motion in several of the bacterial species, namely, \textit{L. cherrii} ($v_{\mathrm{bac}}^{\mathrm{max}} = \SI{15.4}{\micro\meter\per\second}$), \textit{L. dumoffii} ($v_{\mathrm{bac}}^{\mathrm{max}} = \SI{6.1}{\micro\meter\per\second}$), \textit{L. parisiensis} ($v_{\mathrm{bac}}^{\mathrm{max}} = \SI{14.6}{\micro\meter\per\second}$), \textit{P. aeruginosa} ($v_{\mathrm{bac}}^{\mathrm{max}} = \SI{40.1}{\micro\meter\per\second}$), \textit{E. coli} ($v_{\mathrm{bac}}^{\mathrm{max}} = \SI{5.4}{\micro\meter\per\second}$), and \textit{L. pneumophila} conditioned in sterilized tap water for 2 months ($v_{\mathrm{bac}}^{\mathrm{max}} = \SI{13.6}{\micro\meter\per\second}$).
In the case of elongated rod-shaped swimmers (e.g. \textit{L. dumoffii}), the acoustic torque\cite{yamahira2000orientation,schwarz2015rotation} aligned the swimmers parallel to the pressure nodal line.

\section*{Discussion}

Bacteria are known to have a cell volume-dependent volumetric percentage of biomass,\cite{loferer1998determination} which has a density larger than water at $\SI{1.35}{\gram\per\cubic\centi\meter}$ for \textit{E. coli}.\cite{robertson1998determination} This implies that the density of bacterial cells increases with the decrease in the volume of the cells, which is proportional to $\mathrm{ESR}^3$. Since the deviation of the cell density from the density of water increases the acoustic contrast factor, this relation between the cell density and ESR could offer an explanation for the observed dependency of $\Phi_{\mathrm{eff}}$ on the ESR in Fig. \ref{fig:phi_exp_results}(a)-(b) and Fig. \ref{fig:phi_exp_results}(d)-(e). %The results are, however, not conclusive, since the newly discovered relation between the \textit{L. pneumophila} cell concentration and $\Phi_{\mathrm{effM}}$, demonstrated in Fig. \ref{fig:stokesian_results}, could also influence the $\Phi_{\mathrm{effM}}$ reported in Fig. \ref{fig:phi_exp_results}(a).

In the future, it would be important to analyze $\Phi_{\mathrm{eff}}$ of \textit{Legionella} that are grown in amoebae, as the growth process in amoebae has been shown to significantly influence the morphology of the bacteria,\cite{cirillo1994growth} which could influence the acoustic focusability of \textit{Legionella}.

The difference in the acoustic contrast factor of the amoebae in different stages of the encystment cycle could be exploited to acoustically sort the different states of amoebae. This could be realized by combining an on-the-fly analysis of the acoustic contrast factor of an amoeba, similar to the device used by \citet{wang2019continuous}, and a subsequent step that would sort the amoeba into a state-specific outlet based on the readout of the acoustic contrast factor. An applicable acoustic sorting device was already developed, for fluorescent activated sorting of particles.\cite{jakobsson2014acoustic}

Literature suggests that the difference in the size of the two states of \textit{A. castellanii} should be more significant, for example, \citet{bowers1969fine} reported mean diameters of $26.5\pm\SI{0.17}{\micro\meter}$ for trophozoites and $16.2\pm\SI{0.13}{\micro\meter}$ for cysts, but could also be affected by the culturing process.\cite{stratford1978variations} The relatively small sizes of the amoebae that were present in our experiments could stem from the co-culturing with \textit{L. pneumophila}, which are known to destabilize and lyse the host amoebae.\cite{bartram2007legionella} Some of the destabilized amoebae could be lysed during the sample handling steps and during the pumping of the sample into the BAW device; \citet{moffat1992quantitative} for example drew a sample through a 27-gauge needle with an inner diameter of $\SI{210}{\micro\meter}$ three times to lyse the destabilized amoebae; for comparison, the capillaries used to pump-in the sample into our BAW device have an inner diameter of $\SI{100}{\micro\meter}$. During such processes, the larger amoebae would experience higher shear stresses in the narrow sections of the tubing and BAW device, and therefore exhibit a higher probability of lysis, explaining the small size of amoebae in our analysis.

The theoretically demonstrated acceleration of the acoustic focusing and thus higher apparent $\Phi_{\mathrm{eff}}$ explains some previously reported experimental findings. For example, the influence of the \textit{E. coli} concentration on the feasibility of acoustically trapping them has been observed by \citet{hammarstrom2012seed}. Similar concentration dependencies of bacteria-sized objects were observed already before,\cite{miles1995principles} but no study so far, to the best of our knowledge, provided a generally-accepted explanation of the underlying mechanism. Our finding agrees also with a theoretical study in magnetophoresis, where focusing of paramagnetic particles due to a magnetic field was found to accelerate with an increase in particle concentration.\cite{mikkelsen2005microfluidic} We call this phenomenon that stems from the many-body hydrodynamic interactions collective hydrodynamic focusing (CHF).
CHF could apply to many existing studies on acoustophoresis of bacteria and bacteria-sized particles that neglected multi-body effects.\cite{antfolk2014focusing,gerlt2022focusing,devendran2020diffraction,ugawa2022reduced,hawkes1997filtration}
The phenomenon could be significantly influenced by the presence of walls in the experimental devices, analogous to the paradoxical difference in sedimenting particle suspensions of low volumetric concentration that sediment faster when they do not fill the whole container, which allows the fluid flow to go around the cluster of particles.\cite{happel2012low} A similar effect could be caused by local variation in cell concentration or spatial variation of the force field. In addition, specifically to acoustophoresis~\textemdash~the presence of the acoustic microstreaming that can affect the ARF\cite{baasch2019acoustic,pavlic2022shapes} as well as the AIF\cite{pavlic2022aif}~\textemdash~could contribute. These, so-far unexplored influences diminish the CHF in our system for the measurement of $\Phi_{\mathrm{eff}}$, which we showed experimentally. However, in larger channels and with more complex acoustic modes, past experiments\cite{hammarstrom2012seed} already demonstrated the CHF and its potential benefits.
In the future, studying CHF for differently-sized particles and with an extended computational model that accounts for the walls\cite{durlofsky1989dynamic} and other influences could be used to quantify the multi-body effects in past and future experimental studies. Furthermore, CHF could be exploited to control and improve methods that rely on the presence of larger functionalized\cite{lee2019aptamer} or non-functionalized\cite{hammarstrom2012seed} particles to up-concentrate bacteria.

To acoustically focus all the \textit{Legionella} in an at-line system, including the swimmers, the pressure amplitudes of $\gtrsim \SI{0.7}{\mega\pascal}$ would be required, assuming $f=\SI{4}{\mega\hertz}$ and an appropriate focusing length within a device for a given flow rate. In the literature, $p_{\mathrm{a}}$ of above $\SI{1.5}{\mega\pascal}$ have been reported for similar devices and operating conditions,\cite{qiu2022enhancement} making such an application feasible in the near future.

\section*{Methods and theory}

\begin{footnotesize}

\subsection*{Devices and fabrication}
The lab-on-a-chip devices, an example of which is shown in Fig. \ref{fig:device_setup}, were produced through cleanroom processing and feature a three-layered structure made of silicon and glass. A blank silicon wafer ($200\pm\SI{10}{\micro\meter}$ thickness) was first anodically bonded to a glass wafer ($\SI{500}{\micro\meter}$ thickness). The channel designs with $\SI{200}{\micro\meter}$ and $\SI{700}{\micro\meter}$ width of the main channel were then patterned on the silicon through photolithography (resist:  S1828/S1818, Shipley, $4000 \, \mathrm{rpm}$; developer: AZ351B, Microchemicals) and then etched with an inductively coupled plasma deep reactive ion etching (ICP-DRIE) machine (Estrellas, Oxford instruments) through the entire depth of the silicon $\sim\SI{200}{\micro\meter}$.
Afterwards, a glass wafer ($\SI{200}{\micro\meter}$ thickness) was anodically bonded onto the etched silicon-glass wafer. The wafer was then diced into individual $11\times\SI{50}{\milli\meter}$ chips with a wafer saw (DAD3221, Disco corporation). Fused silica capillaries ($164\pm\SI{6}{\micro\meter}$ outer diameter, $100\pm\SI{6}{\micro\meter}$ inner diameter, Molex) were inserted into the inlets and outlets of the chips and fixed with a two-component glue (5 Minute Epoxy, Devcon). A piezoelectric transducer (PZT) (length$\times$width$\times$thickness $=$ $25\times1.55\times\SI{1}{\milli\meter}$ for $\SI{700}{\micro\meter}$-channel width device and $20\times4\times\SI{0.5}{\milli\meter}$ for $\SI{200}{\micro\meter}$-channel width device, Pz26, Meggitt Ferroperm) was glued to each device with a conductive epoxy (H20E, EPO-TEK). PZT was attached to the bottom side of the device that was not facing the microscope lens, in order to allow for the use of a high-magnification (100x) solid immersion lens. Copper cables ($\SI{0.15}{\milli\meter}$ diameter) were connected to the PZT using a conductive silver paste and glued to a device with instant glue for mechanical stability.
To fix the devices under a microscope, a chip-holder with the outer dimensions corresponding to a standard microscope glass slide ($\SI{75}{\milli\meter}\times\SI{25}{\milli\meter}$) was designed and laser-cut from acrylic glass (PMMA).

\subsection*{Sample preparation}
Bacterial strains used in the study: \textit{Legionella pneumophila} Philadelphia-1 (DSM 7513), \textit{Legionella pneumophila subsp. fraseri} (DSM 7514), \textit{Legionella pneumophila subsp. pascullei} (DSM 7515), \textit{Legionella anisa} (DSM 17627), \textit{Legionella cherrii} (DSM 19213), \textit{Legionella micdadei} (DSM 16640), \textit{Legionella dumoffii} (DSM 17625), \textit{Legionella parisiensis} (DSM 19216), \textit{Legionella israelensis} (DSM 19235), \textit{Legionella gormanii} (DSM 16641), \textit{Legionella longbeachae} (DSM 10572), \textit{Pseudomonas aeruginosa} (PAO1), and \textit{Escherichia coli} (ATCC25922).

All \textit{Legionella} spp. were grown in buffered yeast extract (BYE) according to \citet{chatfield2013culturing} overnight at $\SI{37}{\celsius}$. \textit{P. aeruginosa} and \textit{E. coli} were grown in lysogeny broth (LB). For acoustophoresis experiments, $\SI{1}{\milli\liter}$ of overnight culture was used. Cells were washed twice by centrifugation at $10000 \times g$ and resuspension of the pellet in $\SI{1}{\milli\liter}$ autoclaved tap water. Cell concentrations were determined by optical density ($\SI{600}{\nano\meter}$) measurements and Neubauer counting chambers (Bioswisstec). For starvation and conditioning of \textit{Legionella} in BYE and sterile tap water, cells were kept in the media during 5 days or two months before washing, resuspension in sterile tap water, and use in acoustophoresis experiments.

\textit{Acanthamoeba castellanii} was cultured in peptone-yeast-glucose (PYG) medium for 3 days at $\SI{30}{\celsius}$ and resuspended in sterile tap water for acoustophoresis experiments. To induce cyst formation cultures were kept at $\SI{4}{\celsius}$ for 7 days prior to resuspension. Co-cultures with \textit{L. pneumophila} Philadelphia-1 were prepared according to \citet{jarraud2013identification}. Briefly, amoebae were resuspended in ``Pages’ Amoeba Saline Solution'' (PAS) and the concentration was adjusted to $10^5 \mathrm{cells}/\SI{}{\milli\liter}$. The amoebae were incubated at $\SI{30}{\celsius}$ after addition of \textit{L. pneumophila} Philadelphia-1 at a multiplicity of infection $\geq 1$ ($\geq 10^5 \mathrm{cells}/\SI{}{\milli\liter}$). Acoustophoresis experiments were performed with infected amoebae harvested from $\SI{2}{\milli\liter}$ co-culture and resuspended in $\SI{500}{\micro\liter}$ of sterile tap water.

\subsection*{Experimental procedure}
For visual observation of phenomena within the channels of the lab-on-a-chip devices, we used a phase-contrast microscope (Leica) with a camera (Leica MC 120 HD) recording at $30$ frames per second. To visualize and characterize the acoustic field inside a device,\cite{barnkob2010measuring} polystyrene (PS) beads (white $4.97\pm \SI{0.06}{\micro\meter}$ and red $2.21\pm \SI{0.03}{\micro\meter}$, microParticles GmbH) were used. A regular syringe was connected to the inlet capillary of a device, for supplying the particles/cells dispersed in water, as well as for cleaning the chip in-between experiments. A function generator (JDS-2900, JUNTEK) was used to drive the PZT, monitored through an oscilloscope (UTD2025CL, UNI-T).
Devices were additionally characterized by measuring the admittance of the PZT with an impedance analyzer (Sine Phase Z-Check 16777k).

Cleaning in-between experiments with different species was performed by a flushing sequence of $\geq \SI{0.1}{\milli\liter}$ of $\SI{10}{\percent}$ sodium hypochlorite solution, $\geq \SI{0.05}{\milli\liter}$ of $\SI{70}{\percent}$ ethanol solution, and $\geq \SI{0.32}{\milli\liter}$ of sterilized tap water from the same batch as the water used for suspending the cells.

The trajectories of PS particles that were used for estimation of the acoustic pressure amplitude ($p_{\mathrm{a}}$) in the channel were obtained through a particle tracking velocimetry plugin TrackMate\cite{tinevez2017trackmate} in Fiji,\cite{schindelin2012fiji} and post-processed with a custom Matlab\cite{MATLAB:2019b} code that fits the theoretical trajectory from eq. (\ref{eq:theoretTraject}) to an experimental trajectory, using $p_{\mathrm{a}}$ and wavenumber $k$ as the fitting parameters. The trajectories of amoebae were obtained the same way as those of PS particles, while an additional cell-segmentation step was performed for bacteria, using iLastik.\cite{berg2019ilastik} Lastly, the acoustic contrast factor ($\Phi_{\mathrm{eff}}$) of cells was determined by using a modified version of the Matlab code that fits the theoretical trajectory to an experimental cell trajectory, using a measured mean equivalent sphere radius (ESR) of individual species as an input parameter and $\Phi_{\mathrm{eff}}$ as the fitting parameter. Depending on the rate of focusing, the number of frames for the analysis of trajectories of bacteria was reduced by a factor of 5 to 15, to minimize the processing time. To minimize the influence of multi-body interactions, only sufficiently isolated PS particles and amoebae were considered.

The size of individual bacteria species was measured from high magnification (1000x) phase-contrast microscopic images, by measuring the maximum and minimum Feret diameters in a software package Fiji, and using a methodology introduced by \citet{sacca2016simple} to compute the mean cell volume $V$, which was used to extract the ESR, specifically,
\begin{equation}
    \textrm{ESR} = \left( \frac{3 V}{4 \pi} \right)^{1/3} .
\end{equation}
The cell concentration of four bacterial samples was determined using Neubauer counting chambers (Bioswisstec); for the rest of the samples, we determined an average conversion factor $3.255 \cdot 10^6 \SI{}{\milli\liter\per\micro\meter\square}$ that transforms the initial number of bacteria in a $\SI{292}{\micro\meter}\times\SI{164}{\micro\meter}$ video frame into the cell concentration in $\SI{}{\milli\liter}$.

The mean swimming speeds of individual bacterial species were determined by visually identifying swimmers and measuring the distance and time that the swimmers travelled when actively swimming in Fiji.

\subsection*{Acoustic radiation force}

In the underlying theory, basic equations are linearized using the perturbation approach.\cite{bruus2012arf} Accordingly, the physical fields are expanded in a series $\square = \square_0 + \square_1 + \square_2 + \dots$, where $\square$ represents the field, while the subscript denotes the respective order. %We assume the amplitude of the first-order velocity $\vect{v}_1$ to be small with respect to the speed of sound $c_{\mathrm{f}}$ (small Mach number assumption).

The time-averaged ARF on an object in an acoustic field is the mapping of a stress tensor $\tsym{\sigma}$ onto the outward pointing surface normal $ \vect{n}(t)$, integrated over the object's oscillating surface $S(t)$, namely,
\begin{equation}
    \vect{F}_{\mathrm{ARF}} = \avg{\int_{S(t)}\tsym{\sigma} \bcdot \vect{n}(t)\mathrm{d}S}.
    \end{equation}
It has been proven that the second-order expression for the ARF can also be written as \cite{doinikov1994acoustic} 
\begin{equation}
\vect{F}_{\mathrm{ARF}}= \int_{S_0} \bkt{\avg{\tsym{\sigma}_2}-\rhofl \avg{\vect{v}_1 \vect{v}_1}} \bcdot \vect{n}_0 \mathrm{d}S,
\label{eq:ARF}
\end{equation}
with the velocity field $\vect{v}$, and the equilibrium density $\rhofl$. The difference between the mean second-order stress tensor $\avg{\tsym{\sigma}_2}$ and the Reynolds stress $\rho_0 \avg{\vect{v}_1 \vect{v}_1}$ is mapped onto the normal $\vect{n}_0$ pointing out of the arbitrary static surface $S_0$ enclosing the object, and integrated over $S_0$. The mean second-order stress tensor is defined as
\begin{equation}
    \avg{\tsym{\sigma}_2} = - \avg{p_2} \boldsymbol{I} + \eta \left(\grad \avg{\vect{v}_2} + \left( \grad \avg{\vect{v}_2} \right)^T \right),
    \label{eq:sigma2}
\end{equation}
with the pressure field $p$ and the dynamic viscosity $\eta$. The microstreaming and other viscous effects at the second-order are contained in the stress tensor $\avg{\tsym{\sigma}_2}$, while the first-order viscous scattering effects are also part of $\vect{v}_1$.

Equation (\ref{eq:ARF}) reduces to the inviscid ARF in eq. (\ref{eq:arf}) by neglecting the microstreaming velocity $\avg{\vect{v}_2}$, and neglecting $\eta$, which simplifies $\avg{p_2}$, as demonstrated by \citet{bruus2012arf}.
In our system, $f\approx\SI{3.9}{\mega\hertz}$ for bacteria, which corresponds to the viscous boundary layer thickness of $\delta = \sqrt{\eta/\left( \pi \rho_0 f \right)} \approx \SI{0.286}{\micro\meter}$, in water, corresponding to $0.33 \leq \delta / \mathrm{ESR} \leq 0.65$. Recently, it was shown by \citet{pavlic2022shapes} that for a cell-like material at $\delta/\mathrm{ESR} \approx 0.7$, the viscous contributions to the ARF can reach up to $\approx\SI{10}{\percent}$ of the inviscid ARF. For amoebae, the viscous contributions are lower due to the larger size.
In our approach, this contribution is implicitly included in the measured $\Phi_{\mathrm{eff}}$.
The inviscid and viscous acoustic interaction force between bacteria are expected to be small compared to the ARF, except in the vicinity of a pressure node.\cite{pavlic2022aif}

The shape of individual cells, under the condition that it does not deviate considerably from a sphere (as for example \textit{L. dumoffii}), has a negligible viscous and inviscid contribution to the ARF, as long as the actual volume of the cell is taken into account.\cite{pavlic2022shapes} However, the overall dynamics could be influenced by the shape-, orientation-, and configuration-dependent drag.\cite{happel2012low,yamahira2000orientation} Both these effects are inherently accounted for in the measured $\Phi_{\mathrm{eff}}$, as defined in eq. (\ref{legio:eq:PhiEffDef}).
The correction to the drag coefficient $K_{yy}$ from eq. (\ref{eq:FdragKyy}) due to the presence of the walls is for bacteria-sized particles in our channels below $\SI{1}{\percent}$, while for amoebae it could go above $\SI{10}{\percent}$.\cite{baasch2018acoustic} In the measurements, the influence is minimized, since the reduced drag due to the presence of the walls is to a similar degree affecting $p_{\mathrm{a}}$ that is measured by tracking the $\SI{4.97}{\micro\meter}$ polystyrene particles at the same focal plane. The correction to the ARF due to the presence of the walls is, however, negligible.\cite{baasch2020acoustic}

\subsection*{Multi-body computational model}
The multi-body computational model is based on the force-torque (FT) Stokesian dynamics algorithm proposed by \citet{durlofsky1987dynamic}, which computes the velocity of an individual particle based on the positions of all the particles ($\boldsymbol{x}$), their sizes, the properties of the fluid, and the external forces $\boldsymbol{F}_{ext}$ acting on the particles. The change in the position of the particles $\Delta \boldsymbol{x}$ can be computed as\cite{bossis1987self}
\begin{equation}
    \Delta \boldsymbol{x} = \boldsymbol{\mathrm{R}}_{FU}^{-1} \boldsymbol{\cdot} \boldsymbol{F}_{ext} \Delta t + k_{B} T \boldsymbol{\nabla} \boldsymbol{\cdot} \boldsymbol{\mathrm{R}}_{FU}^{-1} \Delta t + \boldsymbol{X}(\Delta t),
    \label{eq:dX}
\end{equation}
with the time step of the numerical integration $\Delta t$, Boltzmann constant $k_{B}$, temperature $T$, random displacement due to Brownian motion $\boldsymbol{X}(\Delta t)$, and with the resistance matrix $\boldsymbol{\mathrm{R}}_{FU}$ that describes the many-body hydrodynamic interactions, as well as the near-field lubrication effects.\cite{durlofsky1987dynamic,bossis1987self,brady1993brownian} We neglect the Brownian motion term in eq. (\ref{eq:dX}) involving $\boldsymbol{\nabla} \boldsymbol{\cdot} \boldsymbol{\mathrm{R}}_{FU}^{-1}$, as we are working with relatively low volumetric concentrations of bacteria. The random displacement $\boldsymbol{X}(\Delta t)$ is the only contribution of the Brownian motion to our multi-body system, and it is computed by enforcing a zero mean value and a covariance given as $\text{cov}\left( \boldsymbol{X}(\Delta t), \boldsymbol{X}(\Delta t) \right) = 2 k_{B} T \boldsymbol{\mathrm{R}}_{FU}^{-1} \Delta t$.
For external forces, we consider the inviscid acoustic radiation force from equation (\ref{eq:arf}),\cite{yosioka1955acoustic} and the inviscid acoustic interaction forces (AIF) from \citet{silva2014acoustic}.
Unless stated differently, we compute and update the many-body hydrodynamic interactions ($\mathrm{M}^{\infty}$ matrix) every $20 \Delta t$, in line with \citet{bossis1987self}.

The computational domain that we consider has a cross-section of $\SI{120}{\micro\meter}\times\SI{120}{\micro\meter}$ and the width of $\SI{200}{\micro\meter}$, corresponding to the channel width in the experimental device. The number of randomly dispersed particles $N_{\mathrm{p}}$ in each simulation is determined by the initial cell concentration $\phi_{\mathrm{c}}$, ranging from $N_{\mathrm{p}} = 288$ for $\phi_{\mathrm{c}} = 1\cdot10^8 \mathrm{cells}/\SI{}{\milli\liter}$ to $N_{\mathrm{p}} = 5760$ for $\phi_{\mathrm{c}} = 2\cdot10^9 \mathrm{cells}/\SI{}{\milli\liter}$. The temperature used in simulations is $\SI{300}{\kelvin}$, unless stated otherwise.

\subsection*{Material properties}
For water, we assume the density of $\SI{1}{\gram\per\cubic\centi\meter}$, the speed of sound of $\SI{1500}{\meter\per\second}$, and the dynamic viscosity of $\SI{1}{\milli\pascal\second}$. For polystyrene (PS) and the associated measurement of $p_{\mathrm{a}}$, we use $\Phi_{\mathrm{eff}} = \Phi^{\mathrm{sph}} = 0.165$, which is based on the material properties reported by \citet{selfridge1985approximate}. The compressibility of a solid object is computed as the inverse of the bulk modulus $K_{\mathrm{obj}}$ through the relation
\begin{equation}
    \kappa_{\mathrm{obj}} = \frac{1}{K_{\mathrm{obj}}} = \frac{1}{\rho_{\mathrm{obj}} \left( c_{\mathrm{P}}^2 - \frac{4}{3} c_{\mathrm{S}}^2 \right)} ,
\end{equation}
with the speed of sound of primary and secondary waves, $c_{\mathrm{P}}$ and $c_{\mathrm{S}}$, respectively. The compressibility of a fluid is computed as $\kappa_0 = 1/(\rho_0 c_0^2)$.
Density of the cytoplasm of a prokaryotic cell for the computational model is estimated to be $\SI{1.156}{\gram\per\cubic\centi\meter}$, based on the mass fractions given by \citet{rojas2020mechanical}, namely $\sim 70\%$ water ($\SI{1}{\gram\per\cubic\centi\meter}$), $\sim 15\%$ proteins ($\SI{1.36}{\gram\per\cubic\centi\meter}$ from \cite{quillin2000accurate}), $\sim 7\%$ nucleic acid ($\SI{1.89}{\gram\per\cubic\centi\meter}$ from \cite{spahn2000method}), and other $\sim 8\%$ comprised of sugars and other molecules (assuming $\SI{1.5}{\gram\per\cubic\centi\meter}$).% This property, however, only influences the AIF.

\end{footnotesize}

\section*{Author Contributions}
All authors designed the research; A.P. designed and built the experimental devices; M.V. prepared all biological samples; A.P. and M.V. conducted the experiments; A.P. developed the code for post-processing the experimental results; A.P. and M.V. post-processed all the experimental results; A.P. established the computational code, carried out the simulations, and analyzed the data; A.P. prepared the initial manuscript draft; all authors contributed to scientific discussion; all authors commented on and contributed to the writing of the manuscript.

\section*{Conflicts of interest}
There are no conflicts to declare.

\section*{Data availability}
Data is available from the corresponding author upon reasonable request.

\section*{Acknowledgements}
We would like to acknowledge funding from InnoSuisse through the contract number 49528.1 for the project \textit{Contactless separation and concentration of Legionella species in drinking water installations}, and the support of the project partners at Georg Fischer JRG AG: Stephan Buerli, Philippe Cachot, Enrico Camelin, Antonio de Agostini, Steffen Lehmann, Hop Nguyen, and Simon Obrist. We thank Mingzhen Fan and Mathias Horn for providing \textit{A. castellanii}. We thank Thierry Baasch for supporting the implementation of the algorithm for computing hydrodynamic multi-body interactions. We thank Bernhard Zybach for building the temperature monitoring system.

%%%END OF MAIN TEXT%%%

%\clearpage

% \nolinenumbers
\balance

%%%REFERENCES%%%
\bibliography{bibliography} %You need to replace "rsc" on this line with the name of your .bib file

\providecommand*{\mcitethebibliography}{\thebibliography}
\csname @ifundefined\endcsname{endmcitethebibliography}
{\let\endmcitethebibliography\endthebibliography}{}
\begin{mcitethebibliography}{70}
\providecommand*{\natexlab}[1]{#1}
\providecommand*{\mciteSetBstSublistMode}[1]{}
\providecommand*{\mciteSetBstMaxWidthForm}[2]{}
\providecommand*{\mciteBstWouldAddEndPuncttrue}
  {\def\EndOfBibitem{\unskip.}}
\providecommand*{\mciteBstWouldAddEndPunctfalse}
  {\let\EndOfBibitem\relax}
\providecommand*{\mciteSetBstMidEndSepPunct}[3]{}
\providecommand*{\mciteSetBstSublistLabelBeginEnd}[3]{}
\providecommand*{\EndOfBibitem}{}
\mciteSetBstSublistMode{f}
\mciteSetBstMaxWidthForm{subitem}
{(\emph{\alph{mcitesubitemcount}})}
\mciteSetBstSublistLabelBeginEnd{\mcitemaxwidthsubitemform\space}
{\relax}{\relax}

\bibitem[Bartram \emph{et~al.}(2007)Bartram, Chartier, Lee, Pond, and
  Surman-Lee]{bartram2007legionella}
J.~Bartram, Y.~Chartier, J.~V. Lee, K.~Pond and S.~Surman-Lee, \emph{Legionella
  and the prevention of legionellosis}, World Health Organization, 2007\relax
\mciteBstWouldAddEndPuncttrue
\mciteSetBstMidEndSepPunct{\mcitedefaultmidpunct}
{\mcitedefaultendpunct}{\mcitedefaultseppunct}\relax
\EndOfBibitem
\bibitem[Van~Kenhove \emph{et~al.}(2019)Van~Kenhove, Dinne, Janssens, and
  Laverge]{van2019overview}
E.~Van~Kenhove, K.~Dinne, A.~Janssens and J.~Laverge, \emph{American Journal of
  Infection Control}, 2019, \textbf{47}, 968--978\relax
\mciteBstWouldAddEndPuncttrue
\mciteSetBstMidEndSepPunct{\mcitedefaultmidpunct}
{\mcitedefaultendpunct}{\mcitedefaultseppunct}\relax
\EndOfBibitem
\bibitem[Chatfield and Cianciotto(2013)]{chatfield2013culturing}
C.~H. Chatfield and N.~P. Cianciotto, \emph{Legionella}, Springer, 2013, pp.
  151--162\relax
\mciteBstWouldAddEndPuncttrue
\mciteSetBstMidEndSepPunct{\mcitedefaultmidpunct}
{\mcitedefaultendpunct}{\mcitedefaultseppunct}\relax
\EndOfBibitem
\bibitem[Bruus \emph{et~al.}(2011)Bruus, Dual, Hawkes, Hill, Laurell, Nilsson,
  Radel, Sadhal, and Wiklund]{bruus2011forthcoming}
H.~Bruus, J.~Dual, J.~Hawkes, M.~Hill, T.~Laurell, J.~Nilsson, S.~Radel,
  S.~Sadhal and M.~Wiklund, \emph{Lab on a Chip}, 2011, \textbf{11},
  3579--3580\relax
\mciteBstWouldAddEndPuncttrue
\mciteSetBstMidEndSepPunct{\mcitedefaultmidpunct}
{\mcitedefaultendpunct}{\mcitedefaultseppunct}\relax
\EndOfBibitem
\bibitem[Undvall \emph{et~al.}(2022)Undvall, Garofalo, Procopio, Qiu, Lenshof,
  Laurell, and Baasch]{undvall2022inertia}
E.~Undvall, F.~Garofalo, G.~Procopio, W.~Qiu, A.~Lenshof, T.~Laurell and
  T.~Baasch, \emph{Physical Review Applied}, 2022, \textbf{17}, 034014\relax
\mciteBstWouldAddEndPuncttrue
\mciteSetBstMidEndSepPunct{\mcitedefaultmidpunct}
{\mcitedefaultendpunct}{\mcitedefaultseppunct}\relax
\EndOfBibitem
\bibitem[Apfel(1982)]{apfel1982acoustic}
R.~Apfel, \emph{The British Journal of Cancer. Supplement}, 1982, \textbf{5},
  140\relax
\mciteBstWouldAddEndPuncttrue
\mciteSetBstMidEndSepPunct{\mcitedefaultmidpunct}
{\mcitedefaultendpunct}{\mcitedefaultseppunct}\relax
\EndOfBibitem
\bibitem[Wiklund(2012)]{wiklund2012acoustofluidics}
M.~Wiklund, \emph{Lab on a Chip}, 2012, \textbf{12}, 2018--2028\relax
\mciteBstWouldAddEndPuncttrue
\mciteSetBstMidEndSepPunct{\mcitedefaultmidpunct}
{\mcitedefaultendpunct}{\mcitedefaultseppunct}\relax
\EndOfBibitem
\bibitem[Goddard \emph{et~al.}(2006)Goddard, Martin, Graves, and
  Kaduchak]{goddard2006ultrasonic}
G.~Goddard, J.~C. Martin, S.~W. Graves and G.~Kaduchak, \emph{Cytometry Part A:
  The Journal of the International Society for Analytical Cytology}, 2006,
  \textbf{69}, 66--74\relax
\mciteBstWouldAddEndPuncttrue
\mciteSetBstMidEndSepPunct{\mcitedefaultmidpunct}
{\mcitedefaultendpunct}{\mcitedefaultseppunct}\relax
\EndOfBibitem
\bibitem[Hartono \emph{et~al.}(2011)Hartono, Liu, Tan, Then, Yung, and
  Lim]{hartono2011chip}
D.~Hartono, Y.~Liu, P.~L. Tan, X.~Y.~S. Then, L.-Y.~L. Yung and K.-M. Lim,
  \emph{Lab on a Chip}, 2011, \textbf{11}, 4072--4080\relax
\mciteBstWouldAddEndPuncttrue
\mciteSetBstMidEndSepPunct{\mcitedefaultmidpunct}
{\mcitedefaultendpunct}{\mcitedefaultseppunct}\relax
\EndOfBibitem
\bibitem[Wang \emph{et~al.}(2018)Wang, Liu, Shin, Chen, Cho, Kim, and
  Han]{wang2018single}
H.~Wang, Z.~Liu, D.~M. Shin, Z.~G. Chen, Y.~Cho, Y.-J. Kim and A.~Han,
  \emph{Microfluidics and Nanofluidics}, 2018, \textbf{22}, 1--7\relax
\mciteBstWouldAddEndPuncttrue
\mciteSetBstMidEndSepPunct{\mcitedefaultmidpunct}
{\mcitedefaultendpunct}{\mcitedefaultseppunct}\relax
\EndOfBibitem
\bibitem[Baasch \emph{et~al.}(2018)Baasch, Reichert, Lak{\"a}mper,
  Vertti-Quintero, Hack, i~Solvas, deMello, Gunawan, and
  Dual]{baasch2018acoustic}
T.~Baasch, P.~Reichert, S.~Lak{\"a}mper, N.~Vertti-Quintero, G.~Hack, X.~C.
  i~Solvas, A.~deMello, R.~Gunawan and J.~Dual, \emph{Biophysical Journal},
  2018, \textbf{115}, 1817--1825\relax
\mciteBstWouldAddEndPuncttrue
\mciteSetBstMidEndSepPunct{\mcitedefaultmidpunct}
{\mcitedefaultendpunct}{\mcitedefaultseppunct}\relax
\EndOfBibitem
\bibitem[Jim{\'e}nez \emph{et~al.}(2022)Jim{\'e}nez, Cabezas, Delay, G{\'o}mez,
  and Camacho]{jimenez2022acoustophoretic}
A.~V. Jim{\'e}nez, D.~C.~O. Cabezas, M.~Delay, I.~G. G{\'o}mez and M.~Camacho,
  \emph{Ultrasound in Medicine \& Biology}, 2022\relax
\mciteBstWouldAddEndPuncttrue
\mciteSetBstMidEndSepPunct{\mcitedefaultmidpunct}
{\mcitedefaultendpunct}{\mcitedefaultseppunct}\relax
\EndOfBibitem
\bibitem[Barnkob \emph{et~al.}(2012)Barnkob, Augustsson, Laurell, and
  Bruus]{barnkob2012acoustic}
R.~Barnkob, P.~Augustsson, T.~Laurell and H.~Bruus, \emph{Physical Review E},
  2012, \textbf{86}, 056307\relax
\mciteBstWouldAddEndPuncttrue
\mciteSetBstMidEndSepPunct{\mcitedefaultmidpunct}
{\mcitedefaultendpunct}{\mcitedefaultseppunct}\relax
\EndOfBibitem
\bibitem[Rodgers(1979)]{rodgers1979ultrastructure}
F.~Rodgers, \emph{Journal of Clinical Pathology}, 1979, \textbf{32},
  1195--1202\relax
\mciteBstWouldAddEndPuncttrue
\mciteSetBstMidEndSepPunct{\mcitedefaultmidpunct}
{\mcitedefaultendpunct}{\mcitedefaultseppunct}\relax
\EndOfBibitem
\bibitem[Hammarstr{\"o}m \emph{et~al.}(2012)Hammarstr{\"o}m, Laurell, and
  Nilsson]{hammarstrom2012seed}
B.~Hammarstr{\"o}m, T.~Laurell and J.~Nilsson, \emph{Lab on a Chip}, 2012,
  \textbf{12}, 4296--4304\relax
\mciteBstWouldAddEndPuncttrue
\mciteSetBstMidEndSepPunct{\mcitedefaultmidpunct}
{\mcitedefaultendpunct}{\mcitedefaultseppunct}\relax
\EndOfBibitem
\bibitem[Habibi and Neild(2019)]{habibi2019sound}
R.~Habibi and A.~Neild, \emph{Lab on a Chip}, 2019, \textbf{19},
  3032--3044\relax
\mciteBstWouldAddEndPuncttrue
\mciteSetBstMidEndSepPunct{\mcitedefaultmidpunct}
{\mcitedefaultendpunct}{\mcitedefaultseppunct}\relax
\EndOfBibitem
\bibitem[Antfolk \emph{et~al.}(2014)Antfolk, Muller, Augustsson, Bruus, and
  Laurell]{antfolk2014focusing}
M.~Antfolk, P.~B. Muller, P.~Augustsson, H.~Bruus and T.~Laurell, \emph{Lab on
  a Chip}, 2014, \textbf{14}, 2791--2799\relax
\mciteBstWouldAddEndPuncttrue
\mciteSetBstMidEndSepPunct{\mcitedefaultmidpunct}
{\mcitedefaultendpunct}{\mcitedefaultseppunct}\relax
\EndOfBibitem
\bibitem[Ugawa \emph{et~al.}(2022)Ugawa, Lee, Baasch, Lee, Kim, Jeong, Choi,
  Sohn, Laurell, Ota,\emph{et~al.}]{ugawa2022reduced}
M.~Ugawa, H.~Lee, T.~Baasch, M.~Lee, S.~Kim, O.~Jeong, Y.-H. Choi, D.~Sohn,
  T.~Laurell, S.~Ota \emph{et~al.}, \emph{Analyst}, 2022\relax
\mciteBstWouldAddEndPuncttrue
\mciteSetBstMidEndSepPunct{\mcitedefaultmidpunct}
{\mcitedefaultendpunct}{\mcitedefaultseppunct}\relax
\EndOfBibitem
\bibitem[Schwarz and Dual(2012)]{schwarz2012ultrasonic}
T.~Schwarz and J.~Dual, AIP Conference Proceedings, 2012, pp. 779--782\relax
\mciteBstWouldAddEndPuncttrue
\mciteSetBstMidEndSepPunct{\mcitedefaultmidpunct}
{\mcitedefaultendpunct}{\mcitedefaultseppunct}\relax
\EndOfBibitem
\bibitem[Guti{\'e}rrez-Ramos \emph{et~al.}(2018)Guti{\'e}rrez-Ramos, Hoyos, and
  Ruiz-Su{\'a}rez]{gutierrez2018induced}
S.~Guti{\'e}rrez-Ramos, M.~Hoyos and J.~Ruiz-Su{\'a}rez, \emph{Scientific
  Reports}, 2018, \textbf{8}, 1--8\relax
\mciteBstWouldAddEndPuncttrue
\mciteSetBstMidEndSepPunct{\mcitedefaultmidpunct}
{\mcitedefaultendpunct}{\mcitedefaultseppunct}\relax
\EndOfBibitem
\bibitem[Zhao \emph{et~al.}(2020)Zhao, Wu, Yang, Wu, Gu, Chen, Ye, Xie, Tian,
  Bachman,\emph{et~al.}]{zhao2020disposable}
S.~Zhao, M.~Wu, S.~Yang, Y.~Wu, Y.~Gu, C.~Chen, J.~Ye, Z.~Xie, Z.~Tian,
  H.~Bachman \emph{et~al.}, \emph{Lab on a Chip}, 2020, \textbf{20},
  1298--1308\relax
\mciteBstWouldAddEndPuncttrue
\mciteSetBstMidEndSepPunct{\mcitedefaultmidpunct}
{\mcitedefaultendpunct}{\mcitedefaultseppunct}\relax
\EndOfBibitem
\bibitem[Jepras \emph{et~al.}(1989)Jepras, Clarke, and
  Coakley]{jepras1989agglutination}
R.~Jepras, D.~Clarke and W.~Coakley, \emph{Journal of Immunological Methods},
  1989, \textbf{120}, 201--205\relax
\mciteBstWouldAddEndPuncttrue
\mciteSetBstMidEndSepPunct{\mcitedefaultmidpunct}
{\mcitedefaultendpunct}{\mcitedefaultseppunct}\relax
\EndOfBibitem
\bibitem[Organization
  \emph{et~al.}(2017)Organization\emph{et~al.}]{world2017guidelines}
W.~H. Organization \emph{et~al.}, \emph{Guidelines for drinking-water quality,
  4th edition, incorporating the 1st addendum}, World Health Organization,
  2017\relax
\mciteBstWouldAddEndPuncttrue
\mciteSetBstMidEndSepPunct{\mcitedefaultmidpunct}
{\mcitedefaultendpunct}{\mcitedefaultseppunct}\relax
\EndOfBibitem
\bibitem[Durlofsky \emph{et~al.}(1987)Durlofsky, Brady, and
  Bossis]{durlofsky1987dynamic}
L.~Durlofsky, J.~F. Brady and G.~Bossis, \emph{Journal of Fluid Mechanics},
  1987, \textbf{180}, 21--49\relax
\mciteBstWouldAddEndPuncttrue
\mciteSetBstMidEndSepPunct{\mcitedefaultmidpunct}
{\mcitedefaultendpunct}{\mcitedefaultseppunct}\relax
\EndOfBibitem
\bibitem[Brady(1993)]{brady1993brownian}
J.~F. Brady, \emph{The Journal of Chemical Physics}, 1993, \textbf{98},
  3335--3341\relax
\mciteBstWouldAddEndPuncttrue
\mciteSetBstMidEndSepPunct{\mcitedefaultmidpunct}
{\mcitedefaultendpunct}{\mcitedefaultseppunct}\relax
\EndOfBibitem
\bibitem[Yosioka and Kawasima(1955)]{yosioka1955acoustic}
K.~Yosioka and Y.~Kawasima, \emph{Acta Acustica united with Acustica}, 1955,
  \textbf{5}, 167--173\relax
\mciteBstWouldAddEndPuncttrue
\mciteSetBstMidEndSepPunct{\mcitedefaultmidpunct}
{\mcitedefaultendpunct}{\mcitedefaultseppunct}\relax
\EndOfBibitem
\bibitem[Silva and Bruus(2014)]{silva2014acoustic}
G.~T. Silva and H.~Bruus, \emph{Physical Review E}, 2014, \textbf{90},
  063007\relax
\mciteBstWouldAddEndPuncttrue
\mciteSetBstMidEndSepPunct{\mcitedefaultmidpunct}
{\mcitedefaultendpunct}{\mcitedefaultseppunct}\relax
\EndOfBibitem
\bibitem[Miles \emph{et~al.}(1995)Miles, Morley, Hudson, and
  Mackey]{miles1995principles}
C.~Miles, M.~Morley, W.~Hudson and B.~Mackey, \emph{Journal of applied
  bacteriology}, 1995, \textbf{78}, 47--54\relax
\mciteBstWouldAddEndPuncttrue
\mciteSetBstMidEndSepPunct{\mcitedefaultmidpunct}
{\mcitedefaultendpunct}{\mcitedefaultseppunct}\relax
\EndOfBibitem
\bibitem[Mikkelsen and Bruus(2005)]{mikkelsen2005microfluidic}
C.~Mikkelsen and H.~Bruus, \emph{Lab on a Chip}, 2005, \textbf{5},
  1293--1297\relax
\mciteBstWouldAddEndPuncttrue
\mciteSetBstMidEndSepPunct{\mcitedefaultmidpunct}
{\mcitedefaultendpunct}{\mcitedefaultseppunct}\relax
\EndOfBibitem
\bibitem[Bruus(2012)]{bruus2012arf}
H.~Bruus, \emph{Lab on a Chip}, 2012, \textbf{12}, 1014--1021\relax
\mciteBstWouldAddEndPuncttrue
\mciteSetBstMidEndSepPunct{\mcitedefaultmidpunct}
{\mcitedefaultendpunct}{\mcitedefaultseppunct}\relax
\EndOfBibitem
\bibitem[Sepehrirahnama \emph{et~al.}(2021)Sepehrirahnama, Oberst, Chiang, and
  Powell]{sepehrirahnama2021acoustic}
S.~Sepehrirahnama, S.~Oberst, Y.~K. Chiang and D.~Powell, \emph{Physical Review
  E}, 2021, \textbf{104}, 065003\relax
\mciteBstWouldAddEndPuncttrue
\mciteSetBstMidEndSepPunct{\mcitedefaultmidpunct}
{\mcitedefaultendpunct}{\mcitedefaultseppunct}\relax
\EndOfBibitem
\bibitem[Wei \emph{et~al.}(2004)Wei, Thiessen, and Marston]{wei2004acoustic}
W.~Wei, D.~B. Thiessen and P.~L. Marston, \emph{The Journal of the Acoustical
  Society of America}, 2004, \textbf{116}, 201--208\relax
\mciteBstWouldAddEndPuncttrue
\mciteSetBstMidEndSepPunct{\mcitedefaultmidpunct}
{\mcitedefaultendpunct}{\mcitedefaultseppunct}\relax
\EndOfBibitem
\bibitem[Selfridge(1985)]{selfridge1985approximate}
A.~R. Selfridge, \emph{IEEE transactions on sonics and ultrasonics}, 1985,
  \textbf{32}, 381--394\relax
\mciteBstWouldAddEndPuncttrue
\mciteSetBstMidEndSepPunct{\mcitedefaultmidpunct}
{\mcitedefaultendpunct}{\mcitedefaultseppunct}\relax
\EndOfBibitem
\bibitem[Pavlic \emph{et~al.}(2022)Pavlic, Nagpure, Ermanni, and
  Dual]{pavlic2022shapes}
A.~Pavlic, P.~Nagpure, L.~Ermanni and J.~Dual, \emph{Physical Review E}, 2022,
  \textbf{106}, 015105\relax
\mciteBstWouldAddEndPuncttrue
\mciteSetBstMidEndSepPunct{\mcitedefaultmidpunct}
{\mcitedefaultendpunct}{\mcitedefaultseppunct}\relax
\EndOfBibitem
\bibitem[Happel and Brenner(2012)]{happel2012low}
J.~Happel and H.~Brenner, \emph{Low Reynolds number hydrodynamics: with special
  applications to particulate media}, Springer Science \& Business Media, 2012,
  vol.~1\relax
\mciteBstWouldAddEndPuncttrue
\mciteSetBstMidEndSepPunct{\mcitedefaultmidpunct}
{\mcitedefaultendpunct}{\mcitedefaultseppunct}\relax
\EndOfBibitem
\bibitem[Barnkob \emph{et~al.}(2010)Barnkob, Augustsson, Laurell, and
  Bruus]{barnkob2010measuring}
R.~Barnkob, P.~Augustsson, T.~Laurell and H.~Bruus, \emph{Lab on a Chip}, 2010,
  \textbf{10}, 563--570\relax
\mciteBstWouldAddEndPuncttrue
\mciteSetBstMidEndSepPunct{\mcitedefaultmidpunct}
{\mcitedefaultendpunct}{\mcitedefaultseppunct}\relax
\EndOfBibitem
\bibitem[Dietersdorfer \emph{et~al.}(2018)Dietersdorfer, Kirschner, Schrammel,
  Ohradanova-Repic, Stockinger, Sommer, Walochnik, and
  Cervero-Arag{\'o}]{dietersdorfer2018starved}
E.~Dietersdorfer, A.~Kirschner, B.~Schrammel, A.~Ohradanova-Repic,
  H.~Stockinger, R.~Sommer, J.~Walochnik and S.~Cervero-Arag{\'o}, \emph{Water
  Research}, 2018, \textbf{141}, 428--438\relax
\mciteBstWouldAddEndPuncttrue
\mciteSetBstMidEndSepPunct{\mcitedefaultmidpunct}
{\mcitedefaultendpunct}{\mcitedefaultseppunct}\relax
\EndOfBibitem
\bibitem[Bowers and Korn(1969)]{bowers1969fine}
B.~Bowers and E.~D. Korn, \emph{The Journal of Cell Biology}, 1969,
  \textbf{41}, 786--805\relax
\mciteBstWouldAddEndPuncttrue
\mciteSetBstMidEndSepPunct{\mcitedefaultmidpunct}
{\mcitedefaultendpunct}{\mcitedefaultseppunct}\relax
\EndOfBibitem
\bibitem[Sun(2005)]{sun2005true}
C.~C. Sun, \emph{Journal of Pharmaceutical Sciences}, 2005, \textbf{94},
  2132--2134\relax
\mciteBstWouldAddEndPuncttrue
\mciteSetBstMidEndSepPunct{\mcitedefaultmidpunct}
{\mcitedefaultendpunct}{\mcitedefaultseppunct}\relax
\EndOfBibitem
\bibitem[Quesada~Cabrera \emph{et~al.}(2011)Quesada~Cabrera, Meersman,
  McMillan, and Dmitriev]{quesada2011nanomechanical}
R.~Quesada~Cabrera, F.~Meersman, P.~F. McMillan and V.~Dmitriev,
  \emph{Biomacromolecules}, 2011, \textbf{12}, 2178--2183\relax
\mciteBstWouldAddEndPuncttrue
\mciteSetBstMidEndSepPunct{\mcitedefaultmidpunct}
{\mcitedefaultendpunct}{\mcitedefaultseppunct}\relax
\EndOfBibitem
\bibitem[Gerlt \emph{et~al.}(2022)Gerlt, Paeckel, Pavlic, Rohner, Poulikakos,
  and Dual]{gerlt2022focusing}
M.~Gerlt, A.~Paeckel, A.~Pavlic, P.~Rohner, D.~Poulikakos and J.~Dual,
  \emph{Physical Review Applied}, 2022, \textbf{17}, 014043\relax
\mciteBstWouldAddEndPuncttrue
\mciteSetBstMidEndSepPunct{\mcitedefaultmidpunct}
{\mcitedefaultendpunct}{\mcitedefaultseppunct}\relax
\EndOfBibitem
\bibitem[Devendran \emph{et~al.}(2020)Devendran, Choi, Han, Ai, Neild, and
  Collins]{devendran2020diffraction}
C.~Devendran, K.~Choi, J.~Han, Y.~Ai, A.~Neild and D.~J. Collins, \emph{Lab on
  a Chip}, 2020, \textbf{20}, 2674--2688\relax
\mciteBstWouldAddEndPuncttrue
\mciteSetBstMidEndSepPunct{\mcitedefaultmidpunct}
{\mcitedefaultendpunct}{\mcitedefaultseppunct}\relax
\EndOfBibitem
\bibitem[Hawkes \emph{et~al.}(1997)Hawkes, Limaye, and
  Coakley]{hawkes1997filtration}
J.~Hawkes, M.~Limaye and W.~Coakley, \emph{Journal of Applied Microbiology},
  1997, \textbf{82}, 39--47\relax
\mciteBstWouldAddEndPuncttrue
\mciteSetBstMidEndSepPunct{\mcitedefaultmidpunct}
{\mcitedefaultendpunct}{\mcitedefaultseppunct}\relax
\EndOfBibitem
\bibitem[Wadhwa and Berg(2021)]{wadhwa2021bacterial}
N.~Wadhwa and H.~C. Berg, \emph{Nature Reviews Microbiology}, 2021,
  1--13\relax
\mciteBstWouldAddEndPuncttrue
\mciteSetBstMidEndSepPunct{\mcitedefaultmidpunct}
{\mcitedefaultendpunct}{\mcitedefaultseppunct}\relax
\EndOfBibitem
\bibitem[Yamahira \emph{et~al.}(2000)Yamahira, Hatanaka, Kuwabara, and
  Asai]{yamahira2000orientation}
S.~Y.~S. Yamahira, S.-i. H. S.-i. Hatanaka, M.~K.~M. Kuwabara and S.~A.~S.
  Asai, \emph{Japanese Journal of Applied Physics}, 2000, \textbf{39},
  3683\relax
\mciteBstWouldAddEndPuncttrue
\mciteSetBstMidEndSepPunct{\mcitedefaultmidpunct}
{\mcitedefaultendpunct}{\mcitedefaultseppunct}\relax
\EndOfBibitem
\bibitem[Schwarz \emph{et~al.}(2015)Schwarz, Hahn, Petit-Pierre, and
  Dual]{schwarz2015rotation}
T.~Schwarz, P.~Hahn, G.~Petit-Pierre and J.~Dual, \emph{Microfluidics and
  Nanofluidics}, 2015, \textbf{18}, 65--79\relax
\mciteBstWouldAddEndPuncttrue
\mciteSetBstMidEndSepPunct{\mcitedefaultmidpunct}
{\mcitedefaultendpunct}{\mcitedefaultseppunct}\relax
\EndOfBibitem
\bibitem[Loferer-Krossbacher \emph{et~al.}(1998)Loferer-Krossbacher, Klima, and
  Psenner]{loferer1998determination}
M.~Loferer-Krossbacher, J.~Klima and R.~Psenner, \emph{Applied and
  Environmental Microbiology}, 1998, \textbf{64}, 688--694\relax
\mciteBstWouldAddEndPuncttrue
\mciteSetBstMidEndSepPunct{\mcitedefaultmidpunct}
{\mcitedefaultendpunct}{\mcitedefaultseppunct}\relax
\EndOfBibitem
\bibitem[Robertson \emph{et~al.}(1998)Robertson, Button, and
  Koch]{robertson1998determination}
B.~Robertson, D.~Button and A.~Koch, \emph{Applied and Environmental
  Microbiology}, 1998, \textbf{64}, 3900--3909\relax
\mciteBstWouldAddEndPuncttrue
\mciteSetBstMidEndSepPunct{\mcitedefaultmidpunct}
{\mcitedefaultendpunct}{\mcitedefaultseppunct}\relax
\EndOfBibitem
\bibitem[Cirillo \emph{et~al.}(1994)Cirillo, Falkow, and
  Tompkins]{cirillo1994growth}
J.~D. Cirillo, S.~Falkow and L.~S. Tompkins, \emph{Infection and Immunity},
  1994, \textbf{62}, 3254--3261\relax
\mciteBstWouldAddEndPuncttrue
\mciteSetBstMidEndSepPunct{\mcitedefaultmidpunct}
{\mcitedefaultendpunct}{\mcitedefaultseppunct}\relax
\EndOfBibitem
\bibitem[Wang \emph{et~al.}(2019)Wang, Liu, Shin, Chen, Cho, Kim, and
  Han]{wang2019continuous}
H.~Wang, Z.~Liu, D.~M. Shin, Z.~G. Chen, Y.~Cho, Y.-J. Kim and A.~Han,
  \emph{Lab on a Chip}, 2019, \textbf{19}, 387--393\relax
\mciteBstWouldAddEndPuncttrue
\mciteSetBstMidEndSepPunct{\mcitedefaultmidpunct}
{\mcitedefaultendpunct}{\mcitedefaultseppunct}\relax
\EndOfBibitem
\bibitem[Jakobsson \emph{et~al.}(2014)Jakobsson, Grenvall, Nordin, Evander, and
  Laurell]{jakobsson2014acoustic}
O.~Jakobsson, C.~Grenvall, M.~Nordin, M.~Evander and T.~Laurell, \emph{Lab on a
  Chip}, 2014, \textbf{14}, 1943--1950\relax
\mciteBstWouldAddEndPuncttrue
\mciteSetBstMidEndSepPunct{\mcitedefaultmidpunct}
{\mcitedefaultendpunct}{\mcitedefaultseppunct}\relax
\EndOfBibitem
\bibitem[Stratford and Griffiths(1978)]{stratford1978variations}
M.~P. Stratford and A.~J. Griffiths, \emph{Microbiology}, 1978, \textbf{108},
  33--37\relax
\mciteBstWouldAddEndPuncttrue
\mciteSetBstMidEndSepPunct{\mcitedefaultmidpunct}
{\mcitedefaultendpunct}{\mcitedefaultseppunct}\relax
\EndOfBibitem
\bibitem[Moffat and Tompkins(1992)]{moffat1992quantitative}
J.~F. Moffat and L.~Tompkins, \emph{Infection and Immunity}, 1992, \textbf{60},
  296--301\relax
\mciteBstWouldAddEndPuncttrue
\mciteSetBstMidEndSepPunct{\mcitedefaultmidpunct}
{\mcitedefaultendpunct}{\mcitedefaultseppunct}\relax
\EndOfBibitem
\bibitem[Baasch \emph{et~al.}(2019)Baasch, Pavlic, and
  Dual]{baasch2019acoustic}
T.~Baasch, A.~Pavlic and J.~Dual, \emph{Physical Review E}, 2019, \textbf{100},
  061102\relax
\mciteBstWouldAddEndPuncttrue
\mciteSetBstMidEndSepPunct{\mcitedefaultmidpunct}
{\mcitedefaultendpunct}{\mcitedefaultseppunct}\relax
\EndOfBibitem
\bibitem[Pavlic \emph{et~al.}(2022)Pavlic, Ermanni, and Dual]{pavlic2022aif}
A.~Pavlic, L.~Ermanni and J.~Dual, \emph{Physical Review E}, 2022,
  \textbf{105}, L053101\relax
\mciteBstWouldAddEndPuncttrue
\mciteSetBstMidEndSepPunct{\mcitedefaultmidpunct}
{\mcitedefaultendpunct}{\mcitedefaultseppunct}\relax
\EndOfBibitem
\bibitem[Durlofsky and Brady(1989)]{durlofsky1989dynamic}
L.~J. Durlofsky and J.~F. Brady, \emph{Journal of Fluid Mechanics}, 1989,
  \textbf{200}, 39--67\relax
\mciteBstWouldAddEndPuncttrue
\mciteSetBstMidEndSepPunct{\mcitedefaultmidpunct}
{\mcitedefaultendpunct}{\mcitedefaultseppunct}\relax
\EndOfBibitem
\bibitem[Lee \emph{et~al.}(2019)Lee, Kim, Shin, Go, Lee, Lee, Kim, and
  Jeong]{lee2019aptamer}
S.~Lee, B.~W. Kim, H.-S. Shin, A.~Go, M.-H. Lee, D.-K. Lee, S.~Kim and O.~C.
  Jeong, \emph{Micromachines}, 2019, \textbf{10}, 770\relax
\mciteBstWouldAddEndPuncttrue
\mciteSetBstMidEndSepPunct{\mcitedefaultmidpunct}
{\mcitedefaultendpunct}{\mcitedefaultseppunct}\relax
\EndOfBibitem
\bibitem[Qiu \emph{et~al.}(2022)Qiu, Baasch, and Laurell]{qiu2022enhancement}
W.~Qiu, T.~Baasch and T.~Laurell, \emph{Physical Review Applied}, 2022,
  \textbf{17}, 044043\relax
\mciteBstWouldAddEndPuncttrue
\mciteSetBstMidEndSepPunct{\mcitedefaultmidpunct}
{\mcitedefaultendpunct}{\mcitedefaultseppunct}\relax
\EndOfBibitem
\bibitem[Jarraud \emph{et~al.}(2013)Jarraud, Descours, Ginevra, Lina, and
  Etienne]{jarraud2013identification}
S.~Jarraud, G.~Descours, C.~Ginevra, G.~Lina and J.~Etienne, \emph{Legionella},
  Springer, 2013, pp. 27--56\relax
\mciteBstWouldAddEndPuncttrue
\mciteSetBstMidEndSepPunct{\mcitedefaultmidpunct}
{\mcitedefaultendpunct}{\mcitedefaultseppunct}\relax
\EndOfBibitem
\bibitem[Tinevez \emph{et~al.}(2017)Tinevez, Perry, Schindelin, Hoopes,
  Reynolds, Laplantine, Bednarek, Shorte, and Eliceiri]{tinevez2017trackmate}
J.-Y. Tinevez, N.~Perry, J.~Schindelin, G.~M. Hoopes, G.~D. Reynolds,
  E.~Laplantine, S.~Y. Bednarek, S.~L. Shorte and K.~W. Eliceiri,
  \emph{Methods}, 2017, \textbf{115}, 80--90\relax
\mciteBstWouldAddEndPuncttrue
\mciteSetBstMidEndSepPunct{\mcitedefaultmidpunct}
{\mcitedefaultendpunct}{\mcitedefaultseppunct}\relax
\EndOfBibitem
\bibitem[Schindelin \emph{et~al.}(2012)Schindelin, Arganda-Carreras, Frise,
  Kaynig, Longair, Pietzsch, Preibisch, Rueden, Saalfeld,
  Schmid,\emph{et~al.}]{schindelin2012fiji}
J.~Schindelin, I.~Arganda-Carreras, E.~Frise, V.~Kaynig, M.~Longair,
  T.~Pietzsch, S.~Preibisch, C.~Rueden, S.~Saalfeld, B.~Schmid \emph{et~al.},
  \emph{Nature Methods}, 2012, \textbf{9}, 676--682\relax
\mciteBstWouldAddEndPuncttrue
\mciteSetBstMidEndSepPunct{\mcitedefaultmidpunct}
{\mcitedefaultendpunct}{\mcitedefaultseppunct}\relax
\EndOfBibitem
\bibitem[Mat(2019)]{MATLAB:2019b}
The Mathworks, Inc., Natick, Massachusetts, \emph{{MATLAB version R2019b}},
  2019\relax
\mciteBstWouldAddEndPuncttrue
\mciteSetBstMidEndSepPunct{\mcitedefaultmidpunct}
{\mcitedefaultendpunct}{\mcitedefaultseppunct}\relax
\EndOfBibitem
\bibitem[Berg \emph{et~al.}(2019)Berg, Kutra, Kroeger, Straehle, Kausler,
  Haubold, Schiegg, Ales, Beier, Rudy,\emph{et~al.}]{berg2019ilastik}
S.~Berg, D.~Kutra, T.~Kroeger, C.~N. Straehle, B.~X. Kausler, C.~Haubold,
  M.~Schiegg, J.~Ales, T.~Beier, M.~Rudy \emph{et~al.}, \emph{Nature Methods},
  2019, \textbf{16}, 1226--1232\relax
\mciteBstWouldAddEndPuncttrue
\mciteSetBstMidEndSepPunct{\mcitedefaultmidpunct}
{\mcitedefaultendpunct}{\mcitedefaultseppunct}\relax
\EndOfBibitem
\bibitem[Sacc{\`a}(2016)]{sacca2016simple}
A.~Sacc{\`a}, \emph{PLoS One}, 2016, \textbf{11}, e0151955\relax
\mciteBstWouldAddEndPuncttrue
\mciteSetBstMidEndSepPunct{\mcitedefaultmidpunct}
{\mcitedefaultendpunct}{\mcitedefaultseppunct}\relax
\EndOfBibitem
\bibitem[Doinikov(1994)]{doinikov1994acoustic}
A.~Doinikov, \emph{Proceedings of the Royal Society of London. Series A:
  Mathematical and Physical Sciences}, 1994, \textbf{447}, 447--466\relax
\mciteBstWouldAddEndPuncttrue
\mciteSetBstMidEndSepPunct{\mcitedefaultmidpunct}
{\mcitedefaultendpunct}{\mcitedefaultseppunct}\relax
\EndOfBibitem
\bibitem[Baasch and Dual(2020)]{baasch2020acoustic}
T.~Baasch and J.~Dual, \emph{Physical Review Applied}, 2020, \textbf{14},
  024052\relax
\mciteBstWouldAddEndPuncttrue
\mciteSetBstMidEndSepPunct{\mcitedefaultmidpunct}
{\mcitedefaultendpunct}{\mcitedefaultseppunct}\relax
\EndOfBibitem
\bibitem[Bossis and Brady(1987)]{bossis1987self}
G.~Bossis and J.~F. Brady, \emph{The Journal of Chemical Physics}, 1987,
  \textbf{87}, 5437--5448\relax
\mciteBstWouldAddEndPuncttrue
\mciteSetBstMidEndSepPunct{\mcitedefaultmidpunct}
{\mcitedefaultendpunct}{\mcitedefaultseppunct}\relax
\EndOfBibitem
\bibitem[Rojas(2020)]{rojas2020mechanical}
E.~R. Rojas, \emph{Physical Microbiology}, Springer, 2020, pp. 1--14\relax
\mciteBstWouldAddEndPuncttrue
\mciteSetBstMidEndSepPunct{\mcitedefaultmidpunct}
{\mcitedefaultendpunct}{\mcitedefaultseppunct}\relax
\EndOfBibitem
\bibitem[Quillin and Matthews(2000)]{quillin2000accurate}
M.~L. Quillin and B.~W. Matthews, \emph{Acta Crystallographica Section D:
  Biological Crystallography}, 2000, \textbf{56}, 791--794\relax
\mciteBstWouldAddEndPuncttrue
\mciteSetBstMidEndSepPunct{\mcitedefaultmidpunct}
{\mcitedefaultendpunct}{\mcitedefaultseppunct}\relax
\EndOfBibitem
\bibitem[Spahn \emph{et~al.}(2000)Spahn, Penczek, Leith, and
  Frank]{spahn2000method}
C.~M. Spahn, P.~A. Penczek, A.~Leith and J.~Frank, \emph{Structure}, 2000,
  \textbf{8}, 937--948\relax
\mciteBstWouldAddEndPuncttrue
\mciteSetBstMidEndSepPunct{\mcitedefaultmidpunct}
{\mcitedefaultendpunct}{\mcitedefaultseppunct}\relax
\EndOfBibitem
\end{mcitethebibliography}
\bibliographystyle{rsc} %the RSC's .bst file

\end{document}

\maketitle

\begin{figure*}
\centering
\includegraphics[scale=0.75]{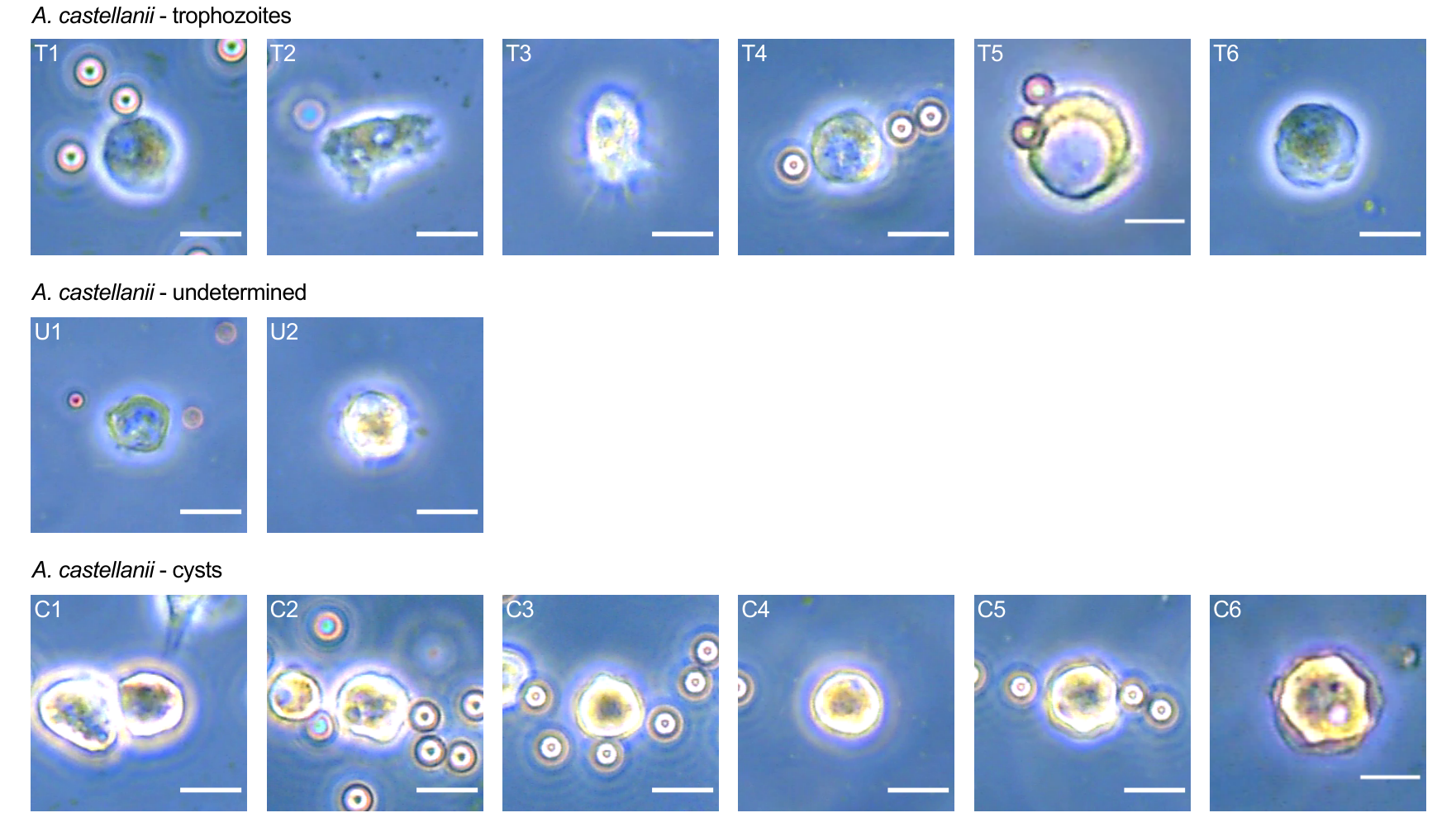}
\caption{Experimentally observed variation in shape and size between investigated \textit{A. castellanii} co-cultured with \textit{L. pneumophila} from Fig. 5 of the main manuscript. Images are obtained by a high-magnification (1000x) phase-contrast microscopy. Red and white spheres next to amoebae are reference polystyrene particles with the diameter of $\SI{2.21}{\micro\meter}$ and $\SI{4.97}{\micro\meter}$, respectively. Scale bar corresponds to $\SI{10}{\micro\meter}$.}
\label{fig:phi_amoeba_size_exp_results}
\end{figure*}

\begin{figure*}
\centering
\includegraphics[scale=0.75]{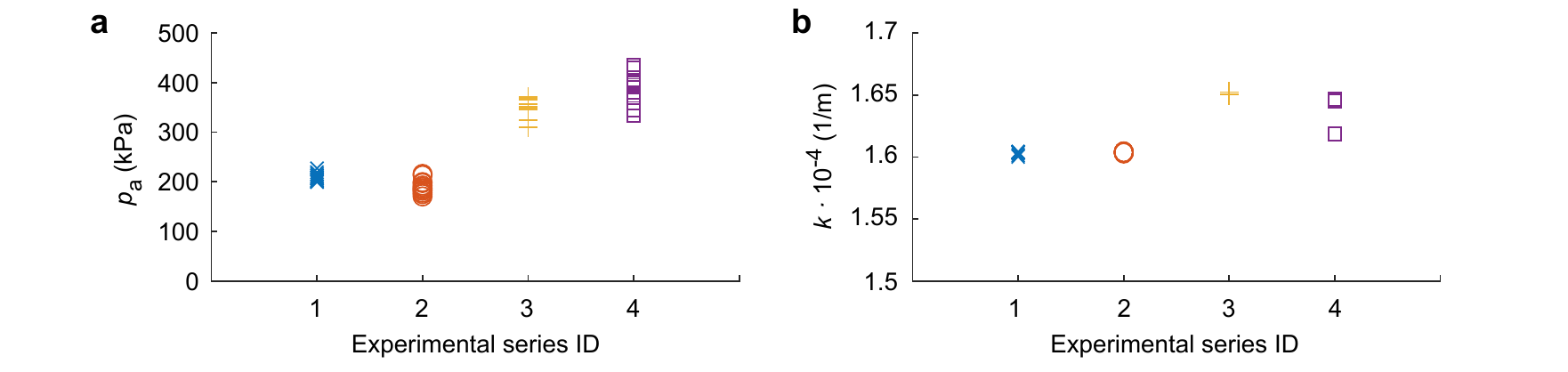}
\caption{Experimentally measured pressure amplitude ($p_{\mathrm{a}}$) and wavenumber ($k$) in experiments series involving bacteria. (a) Pressure amplitude for experimental series 1-3 featured in Fig. 2 and Fig. 4 of the main manuscript, and for experimental series 4 featured in Fig. 6 of the main manuscript. (b) The corresponding wavenumbers. Number of data points for each experimental series: $N = 11$ in series 1, $N=13$ in series 2, $N=14$ in series 3, and $N=14$ in series 4.}
\label{fig:pA_measurement}
\end{figure*}

\begin{figure*}
\centering
\includegraphics[scale=0.75]{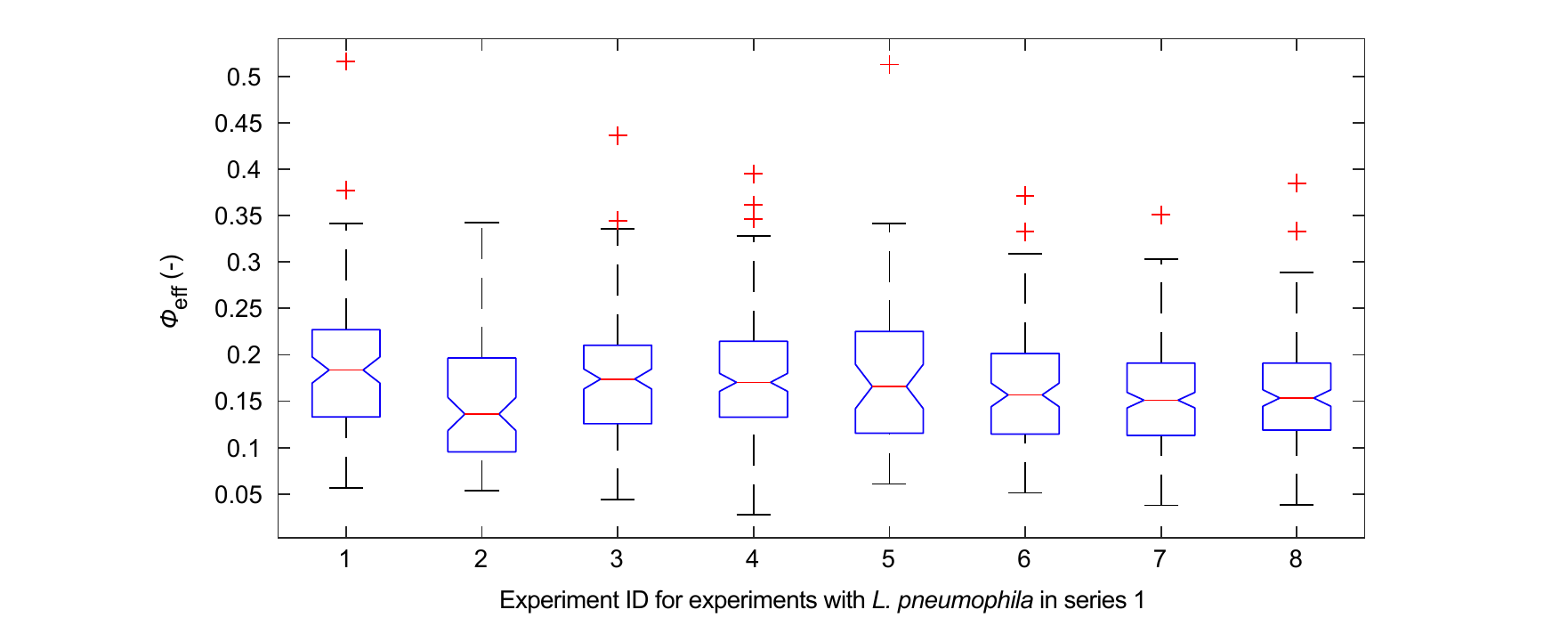}
\caption{Comparison of the measured $\Phi_{\mathrm{eff}}$ of \textit{L. pneumophila} across individual experiments. Red line indicates the median; upper and bottom lines of each box (blue lines), indicate the 75th and 25th percentile, respectively; whiskers indicate the maximum and the minimum; and the red pluses represent the outliers. In Fig. 4 of the main manuscript, all the measurement points from individual experiments of a certain species or sample are combined into a single population.}
\label{fig:statistics_Lpne}
\end{figure*}

\begin{figure*}
\centering
\includegraphics[scale=0.75]{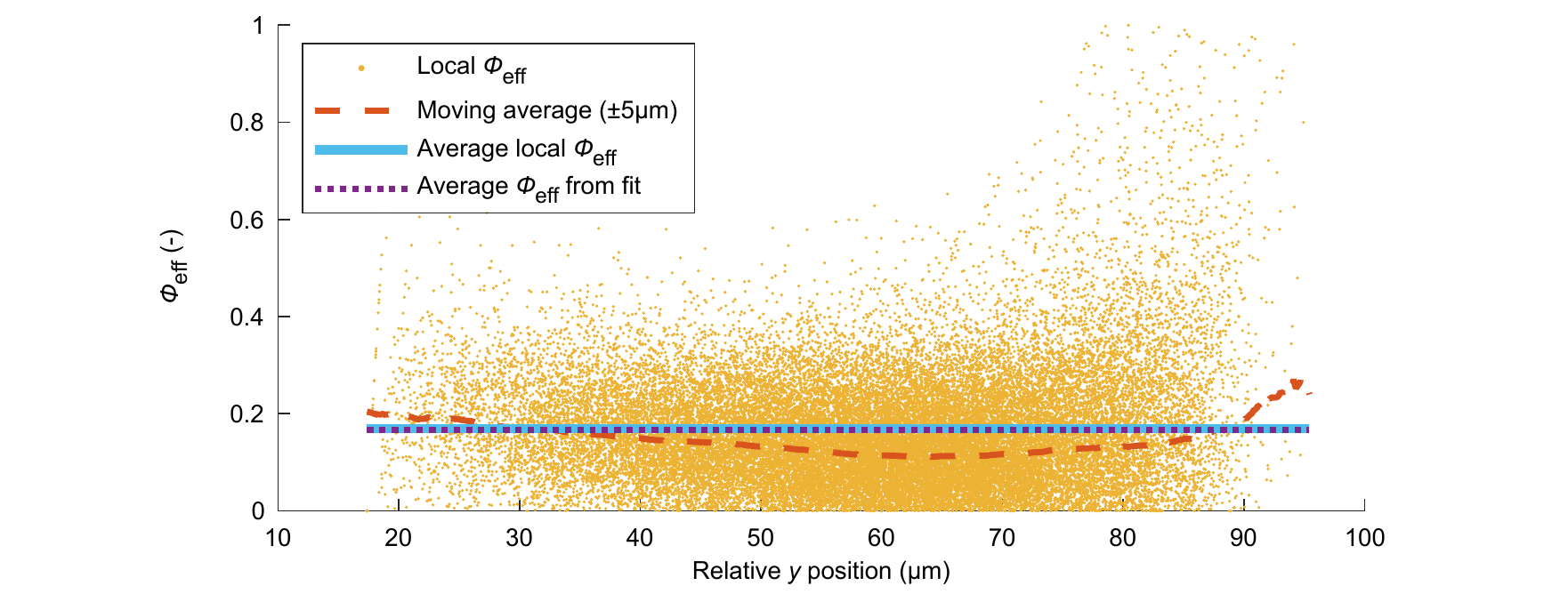}
\caption{Locally measured $\Phi_{\mathrm{eff}}$ (dots) of \textit{L. pneumophila} through eq. (8) of the main manuscript, using a local average velocity $\boldsymbol{v}_{\mathrm{obj}} \boldsymbol{\cdot} \boldsymbol{e}_y$ computed as the path along the $y$-direction in a time interval of $\Delta t = \SI{0.5}{\second}$, divided by $\Delta t$. The locally measured $\Phi_{\mathrm{eff}}$ is represented through a moving average with a window of $\pm \SI{5}{\micro\meter}$ (dashed line). Overall average of the locally measured $\Phi_{\mathrm{eff}}$ (solid line), and the average $\Phi_{\mathrm{eff}}$ (dotted line) computed through fitting the theoretical trajectory to experimental trajectories indicate minimal difference, which can be interpreted as both ways of measuring $\Phi_{\mathrm{eff}}$ giving a comparable average, despite local variations of $\Phi_{\mathrm{eff}}$ along $y$.}
\label{fig:localPhiLpne}
\end{figure*}

\begin{figure*}
\centering
\includegraphics[scale=0.75]{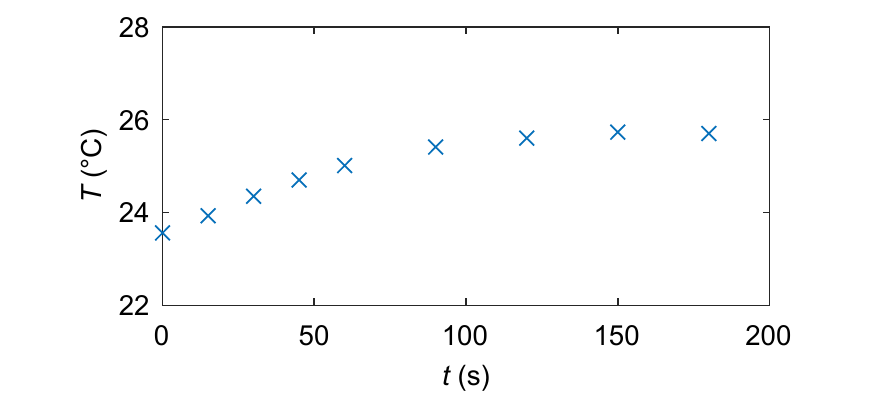}
\caption{Temperature $T$ of the experimental device used for experiments with bacteria ($\SI{200}{\micro\meter}$ channel width) measured during operation under conditions corresponding to experimental series 2 ($f=\SI{3829}{\kilo\hertz}$, $p_{\mathrm{a}} = 190 \pm \SI{13.7}{\kilo\pascal}$). The temperature was measured with a Pt1000 sensor that was connected to the device with a thermal paste, as indicated in Fig. 1(a) of the main manuscript.}
\label{fig:temperature_measurement}
\end{figure*}

\clearpage
%%%REFERENCES%%%
%\bibliography{sample} %You need to replace "rsc" on this line with the name of your .bib file
%\bibliographystyle{rsc} %the RSC's .bst file